
\newif\iffigs\figsfalse              
\figstrue                            
\newif\ifbbB\bbBfalse                
\bbBtrue			     

\input harvmac
\overfullrule=0pt

\def\Title#1#2{\rightline{#1}
\ifx\answ\bigans\nopagenumbers\pageno0\vskip1in%
\baselineskip 15pt plus 1pt minus 1pt
\else
\def\listrefs{\footatend\vskip 1in\immediate\closeout\rfile\writestoppt
\baselineskip=14pt\centerline{{\bf References}}\bigskip{\frenchspacing%
\parindent=20pt\escapechar=` \input
refs.tmp\vfill\eject}\nonfrenchspacing}
\pageno1\vskip.8in\fi \centerline{\titlefont #2}\vskip .5in}

\ifx\answ\bigans\def\tcbreak#1{}\else\def\tcbreak#1{\cr&{#1}}\fi

\newcount\figno \figno=0
\iffigs
\message{If you do not have epsf.tex (to include figures),}
\message{change the option at the top of the tex file.}
\input epsf
\def\fig#1#2#3{\par\begingroup\parindent=0pt\leftskip=1cm\rightskip=1cm
\parindent=0pt \baselineskip=11pt \global\advance\figno by 1 \midinsert
\epsfxsize=#3 \centerline{\epsfbox{#2}} \vskip 12pt
{\bf Fig. \the\figno:} #1\par \endinsert\endgroup\par }
\else
\message{No figures will be included.
 See TeX file for more information.}
\def\fig#1#2#3{\global\advance\figno by 1 \vskip .5in
\centerline{\bf Figure \the\figno} \vskip .5in}
\fi
\def\figlabel#1{\xdef#1{\the\figno}}

\ifbbB
\message{If you do not have msbm (blackboard bold) fonts,}
\message{change the option at the top of the tex file.}
\font\blackboard=msbm10 scaled \magstep1
\font\blackboards=msbm7
\font\blackboardss=msbm5
\newfam\black
\textfont\black=\blackboard
\scriptfont\black=\blackboards
\scriptscriptfont\black=\blackboardss
\def\Bbb#1{{\fam\black\relax#1}}
\else
\def\Bbb{\bf}
\fi

\def\NPB#1#2#3{{\sl Nucl. Phys.} \underbar{B#1} (#2) #3}
\def\PLB#1#2#3{{\sl Phys. Lett.} \underbar{#1B} (#2) #3}
\def\PRL#1#2#3{{\sl Phys. Rev. Lett.} \underbar{#1} (#2) #3}
\def\PRD#1#2#3{{\sl Phys. Rev.} \underbar{D#1} (#2) #3}
\def\CMP#1#2#3{{\sl Comm. Math. Phys.} \underbar{#1} (#2) #3}
\def\hat{\widehat}
\def\tilde{\widetilde}

\def\vev#1{{\langle #1\rangle}}
\def\bo{\hbox{1\kern -.23em {\rm l}}}
\def\bC{{\Bbb C}}

\def\bZ{{\Bbb Z}}
\def\bM{{\bf M}}
\def\CC{{\cal C}}
\def\CF{{\cal F}}
\def\CL{{\cal L}}
\def\CM{{\cal M}}
\def\CN{{\cal N}}
\def\CO{{\cal O}}
\def\CW{{\cal W}}
\def\t{{}^t\!}
\def\p{\partial}
\def\Re{{\rm Re\hskip0.1em}}
\def\Im{{\rm Im\hskip0.1em}}


\Title
{\vbox{\baselineskip12pt\hfill{\vbox{\hbox
{hep-th/9505062, IASSNS-HEP-95/31, RU-95-28}}}}}
{\vbox{\centerline{New Phenomena in $SU(3)$}
\medskip\centerline{Supersymmetric Gauge Theory}}}
\centerline{Philip C. Argyres}
\smallskip
\centerline{\it School of Natural Sciences}
\centerline{\it Institute for Advanced Study}
\centerline{\it Princeton, NJ 08540, USA}
\centerline{\tt argyres@guinness.ias.edu}
\smallskip
\centerline{and}
\smallskip
\centerline{Michael R. Douglas}
\smallskip
\centerline{\it Dept.\ of Physics and Astronomy}
\centerline{\it Rutgers University}
\centerline{\it Piscataway, NJ 08855-0849, USA}
\centerline{\tt mrd@physics.rutgers.edu}
\baselineskip 18pt
\bigskip
We show that
four-dimensional $\CN=2$ supersymmetric $SU(N)$ gauge theory for $N>2$
necessarily contains vacua with
mutually non-local massless dyons, using only analyticity of the
effective action and the weak coupling limit of the moduli space of vacua.
A specific example is the $\bZ_3$ point in the
exact solution for $SU(3)$, and we study its effective Lagrangian.
We propose that the low-energy theory at this point is an
$\CN=2$ superconformal $U(1)$ gauge theory containing both
electrically and magnetically charged massless hypermultiplets.
\Date{May 1995}

\newsec{Introduction}

\nref\sei{N. Seiberg, {\bf hep-th/9411149}, \NPB{435}{1995}{129}, and
	references therein.  For a brief review, see N. Seiberg, {\bf
	hep-th/9408013},{\it The Power of Holomorphy.}}
\nref\SWi{N. Seiberg and E. Witten, {\bf hep-th/9407087},
        \NPB{426}{1994}{19}.}
\nref\AF{P.C. Argyres and A.E. Faraggi, {\bf hep-th/9411057},
	\PRL{73}{1995}{3931}.}
\nref\KLTYi{A. Klemm, W. Lerche, S. Theisen and S. Yankielowicz,
        {\bf hep-th/9411048}, \PLB{344}{1995}{169}.}
\nref\DS{M.R. Douglas and S.H. Shenker, {\bf hep-th/9503163}, {\it
	Dynamics of SU(N) Supersymmetric Gauge Theory.}}
\nref\witten{E. Witten, \PLB{86}{1979}{283}.}
\nref\ss{J.H. Schwarz and A. Sen, {\bf hep-th/9304154}, \NPB{411}{1994}{35}.}
\nref\DSZ{P.A.M. Dirac, {\sl Proc.~R.~Soc.} \underbar{A133} (1931) 60;
	D. Zwanziger, {\sl Phys.~Rev.} \underbar{176} (1968) 1480, 1489;
	J. Schwinger, {\sl Phys.~Rev.} \underbar{144} (1966) 1087;
	\underbar{173} (1968) 1536.}
\nref\yang{T.S. Tu, T.T. Wu and C.N. Yang, {\sl Scientia Sinica}
	\underbar{21} (1978) 317;
	Y. Kazama, C.N. Yang and A.S. Goldhaber, \PRD{15}{1977}{2287}.}
\nref\ASP{P.C. Argyres, R. Plesser and A. Shapere, {\it The Coulomb Phase
       of N=2 Supersymmetric QCD}, IASSNS-HEP-95/32.}
\nref\seione{N. Seiberg, {\bf hep-th/9402044}, \PRD{49}{1994}{6857}.}
\nref\SIi{K. Intriligator and N. Seiberg, {\bf hep-th/9408155}
	\NPB{431}{1994}{551}.}
\nref\Gates{S.J. Gates, Jr., \NPB{238}{1984}{349}.}
\nref\WO{E. Witten and D. Olive, \PLB{78}{1978}{97};
        M. Prasad and C. Sommerfield, \PRL{35}{1975}{760};
        E. Bogomol'nyi, {\sl Sov. J. Nucl. Phys.} \underbar{24} (1976) 449.}
\nref\KLTYii{A. Klemm, W. Lerche, S. Theisen and S. Yankielowicz, {\bf
        hep-th/9412158}, {\it On the Monodromies of N=2 Supersymmetric
        Yang-Mills Theory.}}
\nref\lebo{A. Lebowitz, {\sl Israel J.\  Math.\ }\underbar{12} (1972) 223.}
\nref\Ell{K. Chandrasekharan, {\it Elliptic Functions,} Springer-Verlag 1985.}
\nref\Snr{N. Seiberg, {\bf hep-th/9309335}, \PLB{318}{1993}{469}.}
\nref\kuta{D. Kutasov, {\bf hep-th/9503086}, {\it A Comment on Duality in
	N=1 Supersymmetric Non-Abelian Gauge Theory.}}
\nref\SIii{K. Intriligator and N. Seiberg, {\bf hep-th/9503179}, {\it
	Duality, Monopoles, Dyons, Confinement and Oblique Confinement in
	Supersymmetric $SO(N_c)$ Gauge Theories.}}
\nref\atiyah{M. Atiyah and N. Hitchin, {\it The Geometry and Dynamics
	of Magnetic Monopoles,} Princeton University, 1988.}
\nref\SWii{N. Seiberg and E. Witten, {\bf hep-th/9408099},
        \NPB{431}{1994}{484}.}
\nref\CFT{G. Mack, \CMP{55}{1977}{1}.}
\nref\kutsch{D. Kutasov and A. Schwimmer, {\bf hep-th/9505004}, {\it On
	Duality in Supersymmetric Yang-Mills Theory.}}
%
Over the last year and a half, the work of Seiberg and
collaborators~\refs{\sei} has led to a remarkable variety of exact
results for four-dimensional supersymmetric gauge theories.  Many old
models for physical phenomena, such as the monopole condensation model
of confinement, have explicit realizations in these theories.  Even
more exciting are the new phenomena which have been discovered, such as
duality between $\CN=1$ gauge theories, novel interacting fixed
points, and chiral symmetry breaking by monopole condensation.

An exact low-energy effective Lagrangian for pure $\CN=2$
supersymmetric $SU(2)$ gauge theory was proposed by
Seiberg and Witten \SWi, and
generalized in \refs{\AF,\KLTYi} to $SU(N)$ gauge groups.  $\CN=2$
supersymmetric gauge theory contains a complex adjoint (matrix) scalar
$\phi$, whose expectation value parameterizes physically distinct
vacua.  We refer to the space of vacua as moduli space, and
gauge-invariant coordinates parametrizing it (moduli) are the $N-1$
quantities $\Tr \phi^n$.

Classically, $\phi$ can take any vev satisfying $[\phi,\phi^+]=0$ and
so it can be diagonalized; let its eigenvalues be $\phi_i$.  For
distinct $\phi_i$ this vev breaks $SU(N)$ to the Abelian group
$U(1)^{N-1}$.  The classical analysis is justified for $\phi_i-\phi_j$
large compared to the scale $\Lambda$ where the gauge coupling of the
unbroken quantum theory would become strong, and the low-energy
physics of such vacua is described by weakly coupled $\CN=2$
supersymmetric $U(1)^{N-1}$ gauge theory.  The semiclassical treatment
also predicts the existence of monopoles and dyons, which form
hypermultiplets of $\CN=2$.  This analysis breaks down if any
non-abelian subgroup remains unbroken at the scale $\Lambda$, which
can be arranged by tuning any of the $\phi_i-\phi_j$ to be
$O(\Lambda)$.

In the exact (quantum) solution of these theories, the moduli space
remains the same, and there is a description of the low-energy physics
of each vacuum as a weakly coupled $\CN=2$ supersymmetric $U(1)^{N-1}$
gauge theory, but one not necessarily written in terms of the original
gauge fields.  The essential feature of the quantum theory is that by
tuning a modulus, one can make a monopole or dyon hypermultiplet
massless.  Each such hypermultiplet will become massless on a
submanifold of moduli space of complex codimension one.  By a duality
transformation to the appropriate `magnetic' variables, the low energy
$U(1)^{N-1}$ theory is equivalent to a theory with one massless
`electron' hypermultiplet.  The intersection of two or more of these
submanifolds results in a submanifold of smaller complex dimension
where the massless hypermultiplets are two or more `electrons,' each
charged with respect to a different $U(1)$ factor.

As pointed out in \DS, the $SU(N)$ solutions of \refs{\AF,\KLTYi} have
in addition other vacua where two or more field operators with {\it
mutually non-local charges} become massless.  These fields create the
standard 't Hooft-Polyakov monopoles visible in the semiclassical
limit, and dyons produced from them by shifts of the $\theta$ angle
\witten.

By mutually non-local charges we mean that
 		\eqn\DSZcond{
	h^{(1)} \cdot q^{(2)} - h^{(2)} \cdot q^{(1)} 	\neq 0,
		}
where $h^{(i)}$ and $q^{(i)}$ are the vectors of magnetic and
electric charges of the $i$'th dyon with respect to the $N-1$ $U(1)$
factors.  When this product is non-zero, no duality transformation
will turn the theory into one with only electrically charged
elementary fields; at least one of the elementary fields will have
magnetic charge.  Thus the Lagrangian must simultaneously contain the
standard vector potential coupling locally to electrons, and the dual
vector potential coupling locally to elementary monopoles.  A
manifestly Lorentz invariant Lagrangian of this type is not known, and
indeed the construction of any Lagrangian of this type is fairly
recent \ss.

Several formulations of the quantum mechanics of a finite number of
electrons and monopoles exist, for example in \refs{\DSZ,\yang}.  No
definite reason was found that its physics could not be sensible and
local.  A necessary condition, satisfied here, is that the product
\DSZcond\ is always integral \DSZ.  The new elements here are that the
particles are massless and can be pair produced ad infinitum, so that
one needs a field-theoretic description.  Indeed, whether there is a
`particle' interpretation of the physics is not at all clear, as we
will see.

We propose that these vacua of $SU(N)$ supersymmetric gauge theory
provide `explicit' local realizations of such theories, thus
demonstrating their existence.  Besides the difficulties associated
with mutually non-local gauge charges, the theories will turn out to
be strongly coupled, so the direct definition is not easy to study.
$SU(N)$ gauge theory at these special vacua is convenient for this
purpose, as it contains only these low energy degrees of freedom
together with pure $U(1)$ gauge multiplets.  All other degrees of
freedom have mass $O(\Lambda)$ and decouple in the low energy limit.

Since these vacua occur at singularities of complex codimension one
submanifolds, they are themselves at least of complex codimension two.
Thus the first place one could see them is at isolated points in the
two complex dimensional $SU(3)$ moduli space.  We focus on this theory
because of its simplicity, though when it is easy to do so, we
describe the generalization to the $SU(N)$ theories.

Near the singular vacuum, the theory contains an adjustable second
scale $m<<\Lambda$, and the effective Lagrangian strongly motivates
the claim that between the scales $\Lambda$ and $m$, the theory is at
an RG fixed point.  Exactly at the special vacuum, $m\rightarrow 0$
and the low-energy theory is a fixed point, an $\CN=2$ superconformal
theory.  By taking the limit $\Lambda\rightarrow\infty$, one defines a
superconformal theory without the extra degrees of freedom of $SU(N)$
gauge theory.  Our $SU(3)$ example produces a $\CN=2$ $U(1)$ gauge
theory coupled to three hypermultiplets, and we propose a definition
of the superconformal theory containing only these degrees of freedom.

Another way to single out the special vacuum is to add the superpotential
$\Tr \phi^3$ to the gauge theory, which produces an $\CN=1$ theory
with discrete ground states, two of which are $\CN=1$ deformations of the
$\CN=2$ fixed point theory.

Although we will give strong evidence for our picture of the physics,
since we do not have a complete understanding of the field-theoretic
description, we also discuss possible alternative interpretations in
detail.  Of course one possible interpretation would be that we have found
evidence that the $SU(N)$ solutions of \refs{\AF,\KLTYi} are
incorrect.  We address this possibility by showing that the existence
of vacua with massless dyons with mutually non-local charges follows
solely from analyticity and the topology of the embedding of the
$SU(2)$ moduli space found in \SWi, in the weak coupling limit of the
$SU(3)$ moduli space.  Thus it is assured independent of the details
of the solutions of \refs{\AF,\KLTYi}.  Further confirmation of the
solutions can be found in the physically sensible results and
interpretation of \DS, and by a partially independent derivation, as a
limit of the $N_f=2N_c$ solution found in \ASP.

A competing interpretation of the fixed point would be as an
interacting non-abelian Coulomb phase of the sort found in
\refs{\seione,\SIi}.  The obvious test of this possibility is to look
for non-abelian gauge bosons, which by definition must be present in a
non-abelian Coulomb phase.  Using the solution of \refs{\AF,\KLTYi}, we
will show that they are not.

In section 2 we describe the non-local vacua in the context of a
detailed picture of the complete $SU(3)$ moduli space.  In section 3 we
show that their existence follows from analyticity and the topology of
the embedding of the $SU(2)$ moduli space found in \SWi, in the weak
coupling limit of the $SU(3)$ moduli space.  In section 4 we compute
the effective action near these points, both in the $\CN=2$ theory, and
in the $\CN=1$ theory with a renormalizable superpotential.  In section
5 we compare physical interpretations of the theory, and conclude that
the evidence favors the interpretation as a $U(1)$ theory with mutually
non-local charged fields, and furthermore that the theory at
intermediate scales is a fixed point theory, an interacting $\CN=2,
D=4$ superconformal field theory.  We study its basic properties in
section 6.  As observed in previous work, loop contributions of
particles with $U(1)$ magnetic charges tend to make the electric gauge
coupling relevant.  We point out that given coexisting particles with
mutually non-local charges, there is a novel way to produce fixed
points -- their contributions to the beta function can cancel.  We show
that many features of our effective Lagrangian can be explained by this
interpretation.

\newsec{Singularities in $SU(3)$ Moduli Space}

Gauge-invariant coordinates on the $SU(N)$ moduli space can be taken
to be the elementary symmetric polynomials $s_\ell$, $\ell=2,\ldots,N$,
in the eigenvalues of $\vev{\phi}$
		\eqn\ucoord{
	\det(x-\vev{\phi}) = \sum_{\ell=0}^N (-1)^\ell s_\ell x^{N-\ell}
		}
($s_0=1$ and $s_1$=0 by the $SU(N)$ tracelessness condition).
At a generic point in moduli space where $\vev{\phi}$ breaks the gauge
symmetry to $U(1)^{N-1}$, the low energy effective Lagrangian can be
written in terms of the $\CN=2$ $U(1)$ gauge multiplets $(A_i,W_i)$,
$i=1,\ldots,N-1$, where the $A_i$ are $\CN=1$ chiral superfields and
the $W_i$ are $\CN=1$ (chiral) gauge superfields.  We denote the scalar
component of $A_i$ by $a_i$.  The $\CN=2$ effective Lagrangian is
determined by an analytic prepotential $\CF(A_i)$ \Gates\ and takes the
form
		\eqn\effL{
	\CL_{\rm eff} = {\rm Im}{1\over4\pi} \left[
	\int\!\! d^4\theta\, A_D^i\,{A_i^+} +
	{1\over2}\int\!\!d^2\theta\,\tau^{ij}\,W_i W_j \right] ,
		}
where the dual chiral fields and the effective couplings are given by
		\eqn\defatau{
	A_D^i \equiv {\partial\CF\over\partial A_i},\qquad
	\tau^{ij} \equiv {\partial^2\CF\over\partial A_i\partial A_j} .
		}
Typically, this effective action is good for energies less than
$\Lambda$, the $SU(N)$ strong-coupling scale,  except for regions
of size $\sim\Lambda$ around special submanifolds of moduli space
where extra states become massless.  As we approach these submanifolds
the range of validity of \effL\ shrinks to zero; on these singular
submanifolds the effective Lagrangian must be replaced with one which
includes the new massless degrees of freedom.

\subsec{Charges and Monodromies}

The $U(1)^{N-1}$ effective theory has a lattice of allowed electric and
magnetic charges, $q^i$ and $h_i$ ,
generated by the fundamental representation weights $(q^{(\ell)})^i =
\delta^{\ell,i}$ for the electric charges, and the dual basis
$(h^{(\ell)})_i = \delta^\ell_i$ for the magnetic charge lattice.
BPS saturated hypermultiplets in these theories have effective Lagrangian
		\eqn\BPSeff{\eqalign{
	&\int d^4\theta\ M^+ e^{V\cdot q + V_D \cdot h} M +
			     \tilde M^+ e^{-V\cdot q - V_D \cdot h} \tilde M\cr
	&+ \int d^2\theta\ \sqrt2\ M (A\cdot q + A_D \cdot h)\tilde M + h.c.
		}}
so have a mass given by \WO
		\eqn\BPS{
	M = \sqrt2 |a\cdot q + a_D \cdot h| .
		}
The physics described by the $U(1)^{N-1}$ effective theory is invariant
under an $Sp(2N-2;\bZ)$ group of duality transformations, which acts on
the scalar fields and their duals, as well as the electric and magnetic
charges of all states.  Encircling any singularity in moduli space
(submanifold where extra states become massless) produces a non-trivial
$Sp(2N-2;\bZ)$ transformation.  Thus $\CF(A)$ and the scalar fields
$a_i$ of the effective Lagrangian are not single-valued functions on
the moduli space.

More explicitly, $Sp(2N-2,\bZ)$ consists of all $(2N-2)\times(2N-2)$
integer matrices $\bM$ satisfying $\bM\cdot {\bf I} \cdot\t\bM=\bf I$
where ${\bf I}= {0\,\,\,1\choose-1\,0}$ is the symplectic metric.
The $(2N-2)$--component vector of scalar fields, $\bf a$, as
well as the vector of charges $\bf n$ (thought of as column vectors),
		\eqn\vecs{
	\t{\bf a} \equiv (a_D^i,a_j), \qquad
	\t{\bf n} \equiv (h_i,q^j),
		}
transform under $\bM\in Sp(2N-2,\bZ)$ as
		\eqn\dualtrans{
	{\bf a} \rightarrow \bM\cdot {\bf a}, \qquad
	{\bf n} \rightarrow \t\bM^{-1}\cdot {\bf n}.
		}

In a vacuum with massless charged particles, the $U(1)$'s that couple
to them will flow to zero coupling in the infrared and will be
well-described by perturbation theory.  We use this to compute the
monodromy around a submanifold of such vacua in moduli space.
Consider the case of a submanifold along which one dyon of charge $\bf n$
is massless.  There exists a duality transformation which takes
this to a state with an electric charge $q^1=\gcd(n_i)$
with respect to the
first $U(1)$ factor, say, and otherwise uncharged both electrically and
magnetically.  Thus, in these coordinates, by \BPS, the submanifold along
which this dyon is massless is given locally by $a_1=0$, and the other
$a_i$ vary along this submanifold.
For $a_1 \rightarrow 0$, the leading $a_1$ dependence of the effective
couplings $\tau^{ij}$ is determined to be
		\eqn\taudep{
	\tau^{ij} = \delta^i_1\delta^j_1 {(q^1)^2 \over2\pi i}
	\log(q^1 a_1) + {\cal O}(a_1^0) ,
		}
by a one-loop computation in the effective theory.
Integrating, using \defatau, gives
		\eqn\aDdep{
	a_D^j = \delta^j_1 {(q^1)^2 a_1 \over2\pi i}
        \log(q^1 a_1) + {\cal O}(a_1^0) .
		}
Then, the monodromy $\bM$ around a path $\gamma(t) = \{a_1(\theta) =
e^{i\theta}, a_j=$constant$, j\neq1\}$, which encircles the $a_1=0$
submanifold, is then easily computed from \dualtrans\ to be
                \eqn\monod{
        \bM = \pmatrix{
        \bo&(q^1)^2{\bf e}_{11}\cr
        0&\bo\cr},
		}
where $({\bf e}_{11})^{ij} \equiv \delta^i_1\delta^j_1$.
Converting back to the original description of the physics in which
the charge of the massless dyon was $\bf n$ by the inverse duality
transformation, gives the general form for the monodromy around
a massless dyon singularity to be
                \eqn\dyonmon{
        \bM = \bo+{\bf n}\otimes\t({\bf I}\cdot{\bf n})
	= \pmatrix{ \bo+ q^i h_j & q^i q^j\cr - h_i h_j &
        \bo - h_i q^j\cr}.
		}

The condition on the charges for two dyons to be mutually local is
that they be symplectically orthogonal---{\it c.f.}\ Eq.\ \DSZcond:
		\eqn\mutloc{
	0= \t{\bf n}^{(1)} \cdot {\bf I} \cdot {\bf n}^{(2)}
	= h^{(1)}\cdot q^{(2)} - h^{(2)}\cdot q^{(1)}.
		}
This is equivalent to the condition
that their associated monodromies \dyonmon\ commute:
		\eqn\moncom{
	{}[\bM^{(1)} , \bM^{(2)}] = 0 .
		}
When the
constraint \mutloc\ is satisfied, there exists a symplectic transformation
to dual fields where each dyon is now described as an
electron charged with respect to only one dual low
energy $U(1)$: $h^{(i)} \rightarrow 0$ and $q^{(i)j} \rightarrow
\delta^{i,j} \gcd(n_k^{(i)})$.  Note that there can at most be
$N-1$ linearly independent charge vectors satisfying \mutloc.

\subsec{$SU(N)$ Solution}

The solution \refs{\AF,\KLTYi} for the effective prepotential
$\CF$ is most simply expressed
in terms of an auxiliary Riemann surface $\CC$ which varies over the
moduli space, defined by the curve
		\eqn\curve{\eqalign{
	y^2 &= P(x)^2 - \Lambda^{2N} \cr
	P(x) &\equiv \half\det(x-\vev{\phi})
	= \half\sum_\ell(-1)^\ell s_\ell x^{N-\ell}.\cr
		}}
$\CC$ is a genus $N-1$ Riemann surface realized as a two-sheeted
cover of the complex $x$--plane branched over $2N$ points.  Choose a
basis of $2N-2$ one-cycles $(\alpha_i,\beta_j)$ on $\CC$ with the standard
intersection form $\vev{\alpha_i,\beta_j} = \delta_{ij}$,
$\vev{\alpha_i,\alpha_j} = \vev{\beta_i,\beta_j}=0$.
The $(a_{Di},a_j)$ are then integrals of the meromorphic form
		\eqn\meroform{
	\lambda = {1\over2\pi i}{\p P(x)\over \p x} {x\,dx\over y}
		}
over the $(\alpha_i,\beta_j)$ cycles \AF.  Defining the matrices
		\eqn\ABmat{
	A^i_\ell \equiv {\partial a_D^i\over\partial s_\ell }
	= \oint_{\alpha_i}\lambda_\ell, \qquad
	B_{j\ell} \equiv {\partial a_j\over\partial s_\ell }
	= \oint_{\beta_j}\lambda_\ell,
		}
where $\lambda_\ell=\p\lambda/\p s_\ell$ form a basis of the $N-1$ independent
holomorphic one-forms on $\CC$,
the matrix of $U(1)$ effective couplings is given by
		\eqn\taus{
	\tau^{ij} = \sum_\ell A^i_\ell (B^{-1})^{\ell j},
		}
which is just the period matrix of the Riemann surface $\CC$.

The moduli space contains submanifolds along which a charged particle
becomes massless.  By the mass formula \BPS\ and since the one-form
$\lambda$ \meroform\ is non-singular at the branch points, the only way
this can happen is for cycles of the curve $\CC$ to degenerate.
This occurs whenever two or more of the zeros of the polynomial $Q(x)
\equiv P(x)^2-\Lambda^{2N}$ coincide.  These
degenerations are given by the vanishing of the discriminant of the
polynomial, defined by $\Delta(Q) \equiv \prod_{m>n} (e_m-e_n)^2$,
where the $e_m$ are the $2N$ zeros of $Q(x)$.  $Q$ factorizes as $Q=Q_+
Q_-$ with
		\eqn\factors{
	Q_\pm(x) = P(x) \pm \Lambda^N .
		}
Denote the zeros of $Q_\pm$ by $e_i^\pm$.  Then, from its definition,
the discriminant of $Q$ is $\Delta(Q) = \Delta(Q_+) \Delta(Q_-)
\prod_{i,j} (e^+_i-e^-_j)^2 = \Delta(Q_+) \Delta(Q_-) \prod_i
Q_+(e_i^-)^2$.  Since $Q_+ - Q_- = 2\Lambda^N$, $Q_+(e_i^-) =
2\Lambda^N$, so the discriminant factorizes
		\eqn\facdisct{
	\Delta(Q) = 2^{2N} \Lambda^{2N^2} \Delta(Q_+) \Delta(Q_-).
		}
Thus there are always two separate codimension one singular submanifolds
in moduli space, described by the vanshing of $\Delta(Q_\pm)$.  As we will
see later, these two submanifolds each correspond to one of
the two singular points in the $SU(2)$ moduli space, which is embedded in
a complicated way in the $SU(N)$ moduli space at weak coupling.

\subsec{$SU(3)$ Moduli Space}

We now specialize to the $SU(3)$ case, and denote the good global
coordinates on its moduli space by
		\eqn\notn{
	u \equiv -s_2 = -\phi_1\phi_2-\phi_1\phi_3-\phi_2\phi_3,
	\qquad v \equiv s_3 =\phi_1\phi_2\phi_3.
		}
Note that there is a $\bZ_6$ spontaneously broken discrete global
symmetry whose action on the $SU(3)$ moduli space is generated by
$u\rightarrow e^{2\pi i/3}u$, $v\rightarrow e^{i\pi} v$.
We find from \curve
		\eqn\discpm{
	\Delta(Q_\pm) = 4 u^3 - 27 (v \pm 2\Lambda^3)^2 .
		}
The two submanifolds $\Delta(Q_\pm) =0$ intersect at the three points $v=0$,
$u^3 = (3\Lambda^2)^3$.  These points, which we refer to
as the `$\bZ_2$ vacua' since each leaves a $\bZ_2\subset\bZ_6$ unbroken,
correspond to vacua where two mutually local
dyons are simultaneously massless.  Their physics was studied in detail in
\refs{\AF,\DS}.  Upon breaking $\CN=2$ to $\CN=1$ supersymmetry
by relevant or marginal terms in the superpotential, these
points are not lifted, and so correspond to the
three discrete vacua of the $\CN=1$ $SU(3)$ theory.

In addition to these these intersection points, there are also
singular points of each curve individually: {\it e.g.}\ points
at which $\partial\Delta(Q_+)/\partial u = \partial\Delta(Q_+)/\partial
v = 0$.  There is one such point on the $\Delta(Q_+)$ curve:
$u=0$, $v=-2\Lambda^3$.  A similar point with $v \rightarrow -v$
occurs on the $\Delta(Q_-)$ curve.  These points, which we refer to
as the `$\bZ_3$ vacua' since they leave unbroken a $\bZ_3\subset\bZ_6$,
will be examined in detail below.

In the remainder of this section, we build up a picture of
the $SU(3)$ moduli space and how the $\Delta(Q_\pm)=0$ curves
sit in it.  We do this by presenting three-dimensional
slices of the four-(real)-dimensional $SU(3)$ moduli space.
This moduli space is  $\bC^2$, parametrized by the two complex
coordinates $u$, $v$.  One nice slice \KLTYii\ of this space is the
hypersurface $\Im v=0$, shown in Fig.\ 1.

\fig{The Im$v=0$ hypersurface in $SU(3)$ moduli space.  The heavy
solid curves are its intersection with the surface of massless dyons
$\Delta(Q_-)=0$, and the dotted curve with the massless dyon surface
$\Delta(Q_+)=0$.  The solid circles mark the
$\bZ_2$ vacua, while the open circles denote the $\bZ_3$ vacua.}
{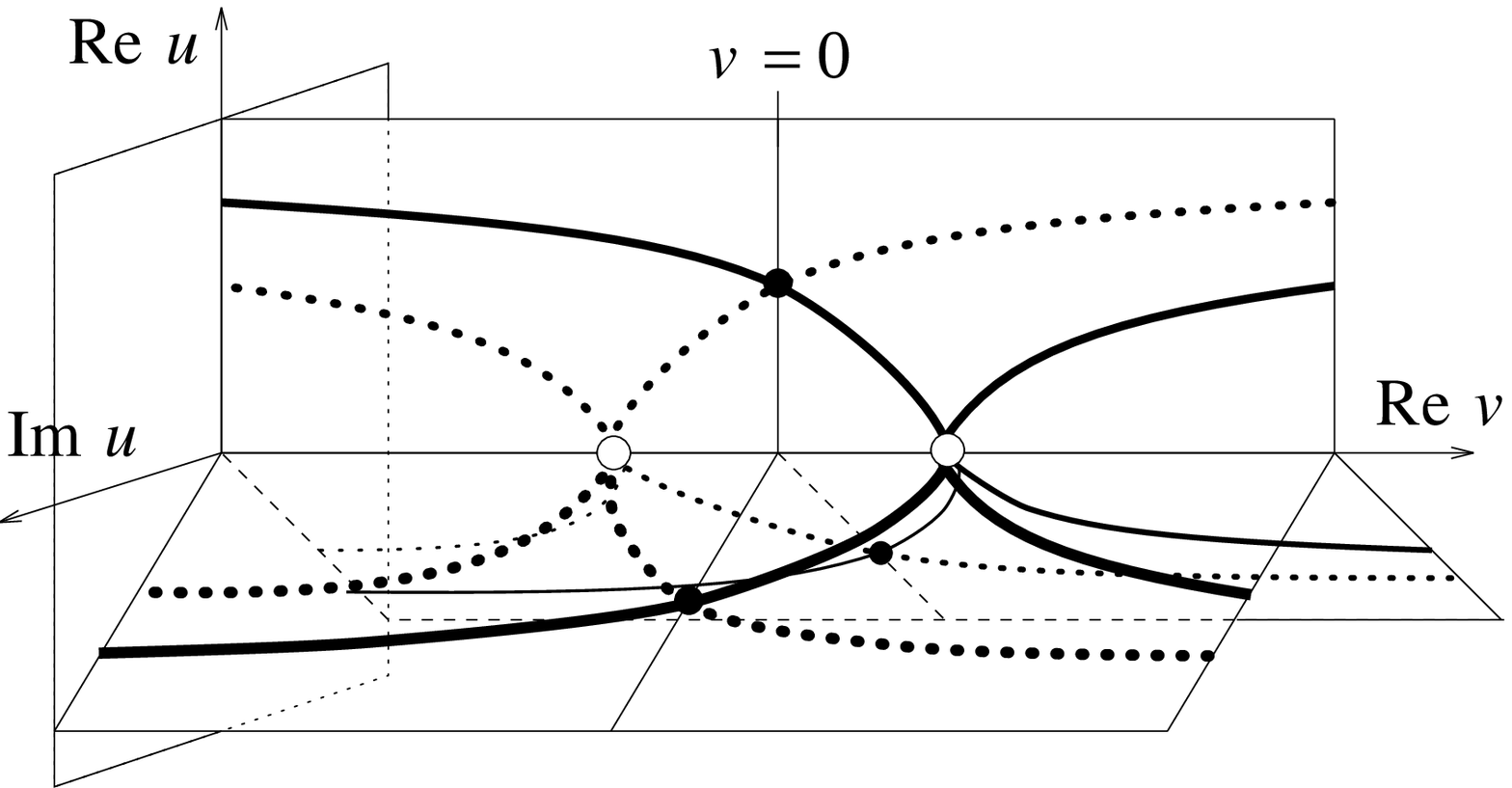}{8.5cm}

The $\bZ_2$ points are seen to correspond to the transverse intersection
of two one-complex-dimensional surfaces in $\bC^2$, as expected.
The nature of the $\bZ_3$ points is less clear, however.
A better understanding of the $\bZ_3$ points can be obtained by
slicing the moduli space by a 3-sphere surrounding one of these
points.  The $\Delta(Q_\pm)$ curves \discpm\ near these
points take a simple form upon shifting
$\tilde v \equiv v\pm 2\Lambda^3$:
		\eqn\Ztform{
	4 u^3 = 27 \tilde v^2 .
		}
Consider the intersection of this surface
with the hypersurface $4|u|^3 +27|\tilde v|^2 = R^6$ which is
topologically a 3-sphere.
The norm of \Ztform\ implies that $4|u|^3 =27|\tilde v|^2 =  \half R^6$,
leaving the torus of phases of $\psi_u\equiv\arg u$ and
$\psi_v\equiv\arg\tilde v$ unconstrained.
The argument of \Ztform\ implies $3\psi_u=2\psi_v \pmod{2\pi}$, whose
solution is a
curve which winds three times around the torus in one direction while it
winds twice in the other---the knot shown in Fig.\ 2.

\fig{The heavy lines are the stereographic projection of the
intersection of a 3-sphere centered on a $\bZ_3$ point with the
corresponding massless dyon curve $\Delta(Q_\pm)=0$.
The lighter curves are three convenient paths encircling the knot.}
{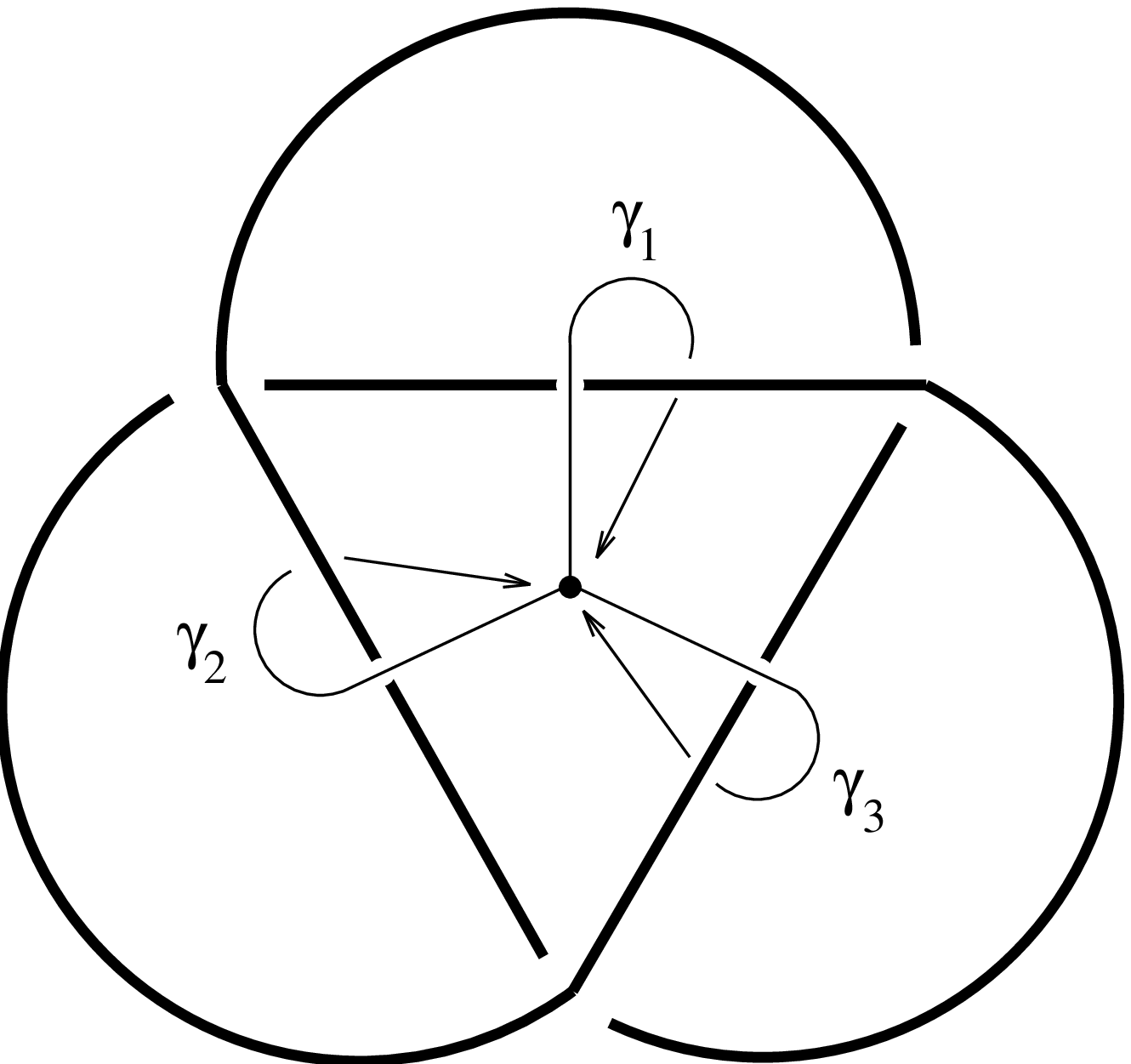}{6cm}

There are in principle two independent $Sp(4,\bZ)$ monodromies that could
occur along paths around such a knot.  This can be seen as follows.
Consider the monodromies $\bM_i$ around the three paths $\gamma_i$
indicated in Fig.\ 2.  Deform $\gamma_1$ by sliding it along the knot, to
become $\gamma_3$ except for its wrapping around the part of the knot that
$\gamma_2$ encircles.  Thus $\gamma_1 \simeq \gamma_2 \cdot \gamma_3
\cdot \gamma_2^{-1}$, as well as cyclic permutations.  This implies the
corresponding relations for their monodromies
		\eqn\monrel{
	\bM_i \bM_{i+1} = \bM_{i+1} \bM_{i+2} ,
		}
which in turn imply that all the monodromies around the knot can be
generated by just two monodromies, say $\bM_1$ and $\bM_2$,
which are constrained to satisfy $\bM_1\bM_2\bM_1 =
\bM_2\bM_1\bM_2$.

Note that either the resulting monodromies do not
commute or they are the
same.  Which of these two possibilities occurs
is a matter of computation of the monodromies realized by the actual
degeneration of the Riemann surface $\CC$ \curve\ near a $\bZ_3$ point.
It turns out that $\bM_1\neq \bM_2$ and thus that
the dyons going massless at the $\bZ_3$ points are indeed mutually
nonlocal.  One easy way of seeing this is to note that the
monodromies around the $\Delta(Q_\pm)=0$ curves have been computed
in Ref.~\KLTYii\ at weak coupling far from the $\bZ_3$ points (at
large $\Re v$ in Fig.\ 1), and that there is no obstruction to deforming
their defining paths down to the $\bZ_3$ point.  With the notation
${\bf n}^{(i)} = (h^{(i)}_1, h^{(i)}_2 ; q^{(i)}_1 , q^{(i)}_2 )$,
the charges of the massless dyons corresponding to shrinking the
$\gamma_i$ were found to be \KLTYii
		\eqn\chrgs{
	\matrix{{\bf n}^{(1)}&=&(&1,&0;&1,&0&),\cr
	{\bf n}^{(2)}&=&(&0,&1;&-1,&1&),\cr
	{\bf n}^{(3)}&=&(&1,&1;&0,&1&),\cr}
		}
which are indeed not mutually local (and only two are linearly
independent).  Alternatively, one can calculate the monodromies
directly from the curve \curve\ near a $\bZ_3$ point to find the same
answer (up to a duality basis transformation).

There is a symplectic transformation to a basis in which the
charges are
		\eqn\chrgsii{
	\matrix{{\bf n}^{(1)}&=&(&1,&0;&0,&0&),\cr
	{\bf n}^{(2)}&=&(&0,&0;&-1,&0&),\cr
	{\bf n}^{(3)}&=&(&1,&0;&-1,&0&).\cr}
		}
In this basis we have one electron, one monopole, and one dyon, all
charged with respect to only the first $U(1)$ factor.

\subsec{Generalization to $SU(N)$}

The topology of the singular submanifolds of the $SU(N)$ moduli spaces
for $N>3$ is much more complicated and harder to analyze.  For example,
the $SU(4)$ moduli space is 3-dimensional (we use only complex
dimensions in this paragraph) and there are still two 2-dimensional
submanifolds $\Delta(Q_\pm)=0$ where a single dyon becomes massless.
These can intersect along 1-dimensional submanifolds where two
mutually-local dyons become massless.  Also, there will be
1-dimensional submanifolds of transverse self-intersections of the
$\Delta(Q_\pm)=0$ curves (where again two mutually local dyons must
become massless---see the discussion in the next section).  The
0-dimensional manifolds where these self-intersections intersect the
other $\Delta(Q_\pm)=0$ curve are a set of four isolated points where
the maximal number of three mutually local dyons are massless.  These
points, and their $SU(N)$ generalizations, were described in \DS.

Since at weak coupling the $SU(3)$ moduli space $\CM_3$ is locally
embedded in the $SU(4)$ moduli space as $\bC\times\CM_3$, the $\bZ_3$
vacua of $\CM_3$ will give rise to a whole 1-dimensional curve of such
singularities.  This curve itself has a cusp-like singularity.  Such
extra-cuspy points are in some sense the $SU(4)$ analogs of the $\bZ_3$
points for $SU(3)$.

The simplest examples are the $SU(N)$ points $P(x)=\half x^N\pm
\Lambda^N$.  At these points, at least $N-1$ mutually non-local dyons
with linearly independent charges become massless.  They are charged
with respect to half (for $N$ odd, or $N/2$ for $N$ even) of the $U(1)$
factors.

\newsec{Existence of Nonlocal Points}

In this section we show that the combination of the Seiberg--Witten
solution \SWi\ of $\CN=2$ $SU(2)$ gauge theory, and the way the $SU(2)$
moduli space is embedded in the $SU(3)$ moduli space at weak coupling
implies that some sort of vacua with
massless mutually non-local dyons (like the $\bZ_3$ vacua) must exist.

The strategy for the argument will be to consider the intersection of
$SU(3)$ moduli space with a large 3-sphere where perturbation theory
reliably computes the embedding of the $SU(2)$ moduli space in the
$SU(3)$ moduli space.  Using the Seiberg-Witten solution then gives the
topology and monodromies of the one-real-dimensional singular curves on
the 3-sphere where a dyon becomes massless.  We then shrink the
3-sphere, sweeping out the whole of the $SU(3)$ moduli space.  The
singular curves will deform continuously along this family of
3-spheres, except for isolated points where the curves may cross (or
small loops may shrink to a point and disappear).  By continuity, the
monodromies around the curves remain unchanged since $Sp(4,\bZ)$ is
discrete.  This is true even when there is a crossing of curves, since
the monodromy can be measured along a path that loops around the curve
anywhere along its length, while the crossing takes place at a single
point.

As we show below, the singular curves on the large 3-sphere are
knotted and have non-commuting monodromies around different sections of
the same curve, just as in the case of the knotted curve in Fig.\ 2.
Now, imagine that we shrink the large 3-sphere to a point in moduli space
that does not lie on any singular surface.  Then the curves must
eventually shrink to points until they have entirely disappeared (since a
sufficiently small 3-sphere about the final point does not intersect any
singular curve, by construction).  But since there are non-commuting
monodromies about different segments of a single curve, by continuity
there must be at least one crossing or degeneration where the parts of the
curve with non-commuting monodromies meet.  Such a point corresponds to a
vacuum with massless mutually non-local dyons, whose existence we sought
to prove.

This argument assumes that the large three-sphere is contractible in
moduli space.  Perhaps the correct moduli space is not $\bC^2$, but
some topologically non-trivial space?
The simplest example would
be $\bC^2$ minus the points with mutually non-local massless particles.
In other words, perhaps the $\bZ_3$ points in the $SU(3)$ moduli
space are really `points at infinity' and so should not be considered
consistent vacua.  However, there will be vacua with
mutually non-local particles as light as one
likes, and one is still faced with the problem of making sense of
such a theory.  Indeed, this will be the gist of our analysis in
later sections:  to understand what constraints there are on
the possible physics at the $\bZ_3$ points by looking at the
$U(1)\times U(1)$ vacua arbitrarily close to them.

One might also imagine that the moduli space has a more complicated
topology, presumably a branched cover of $\bC^2$.  One would need to
propose a physical interpretation of the branch points and, since each
of the covering sheets has a semiclassical limit, identify the
observable distinguishing them in this limit.  We did not find a
scenario of this sort.

We now proceed to the determination of the topology and monodromies of the
singular curves on the large 3-sphere.  Afterwards we include a discussion
designed to make the topological aspects of the above argument more
concrete.

\subsec{$SU(2)$ in $SU(3)$ at Weak Coupling}

Classically, the singularities which reach out to
infinity in $SU(3)$ moduli space are those where $\vev{\phi}$
breaks $SU(3)\rightarrow U(1)\times SU(2)$.  This occurs whenever two
$\vev{\phi}$ eigenvalues are the same, which,
in gauge invariant coordinates, is the
curve
		\eqn\suiiform{
	4u^3 = 27 v^2 .
		}
As this is the same as the curve
\Ztform\ found  near the $\bZ_3$ vacua, its intersection with a 3-sphere
centered on $u=v=0$ will be the same knot, shown in Fig.\ 2.

Quantum corrections modify this classical picture qualitatively, since
near the $SU(2)$ singularity \suiiform\ the low-energy physics is strongly
coupled.  Parametrize the eigenvalues of $\vev\phi$ as $\{ M+a ,
M-a , -2M \}$, implying
		\eqn\uvparam{
	u = 3 M^2 + a^2 , \qquad
	v =-2 M ( M^2 - a^2 ) .
		}
For $|M| \gg |a|$, this vev breaks $SU(3)$ in two stages: $SU(3)
{\buildrel M \over \rightarrow} U(1) \times [ SU(2) {\buildrel a \over
\rightarrow} U(1) ]$.   For $|M| \gg |\Lambda|$, the first $U(1)$ factor
decouples and the low-energy physics is effectively described by an $\CN
= 2$ $SU(2)$ gauge theory spontaneously broken by the adjoint scalar vev
$\vev\phi \sim {a\,\,\,0\choose0\,-a}$.  This theory was solved by
Seiberg and Witten \SWi\ who found that instead of a single singularity at
$a^2=0$ (where classically $SU(2)$ would be restored), there are two
singularities at $a^2 = \pm \tilde\Lambda^2$ where certain dyons are
massless.  Here $\tilde\Lambda$ is the $SU(2)$ strong-coupling scale.
Thus, the single classical curve \suiiform\ is split quantum-mechanically
into two curves.  In terms of its intersection with a large 3-sphere, this
means that the single knot of Fig.\ 2 is split into two linked knots.

The precise way these two knots are linked is fixed by the relation of
$\tilde\Lambda$ to $\Lambda$, determined by the renormalization group
matching $\Lambda^3 \sim M \tilde \Lambda^2$.  This implies that the
$SU(2)$ dyon singularities occur at $a^2=\pm\Lambda^3/M$.  Plugging into
\uvparam\ and eliminating $M$ gives the equations for the singular curves
		\eqn\linkform{
	4u^3 = 27 ( v^2 \mp 8 v \Lambda^3 ) + \CO(\Lambda^6).
		}
The intersection of these two curves with a 3-sphere of radius much
larger than $|\Lambda|$ gives the two knots linked as shown in Fig.\ 3.

\fig{The heavy solid and dotted lines are the stereographic projection of the
intersection of a large 3-sphere with the singular curves at weak coupling
in the $SU(3)$ moduli space.  The lighter curves are
six convenient paths encircling the knots.}
{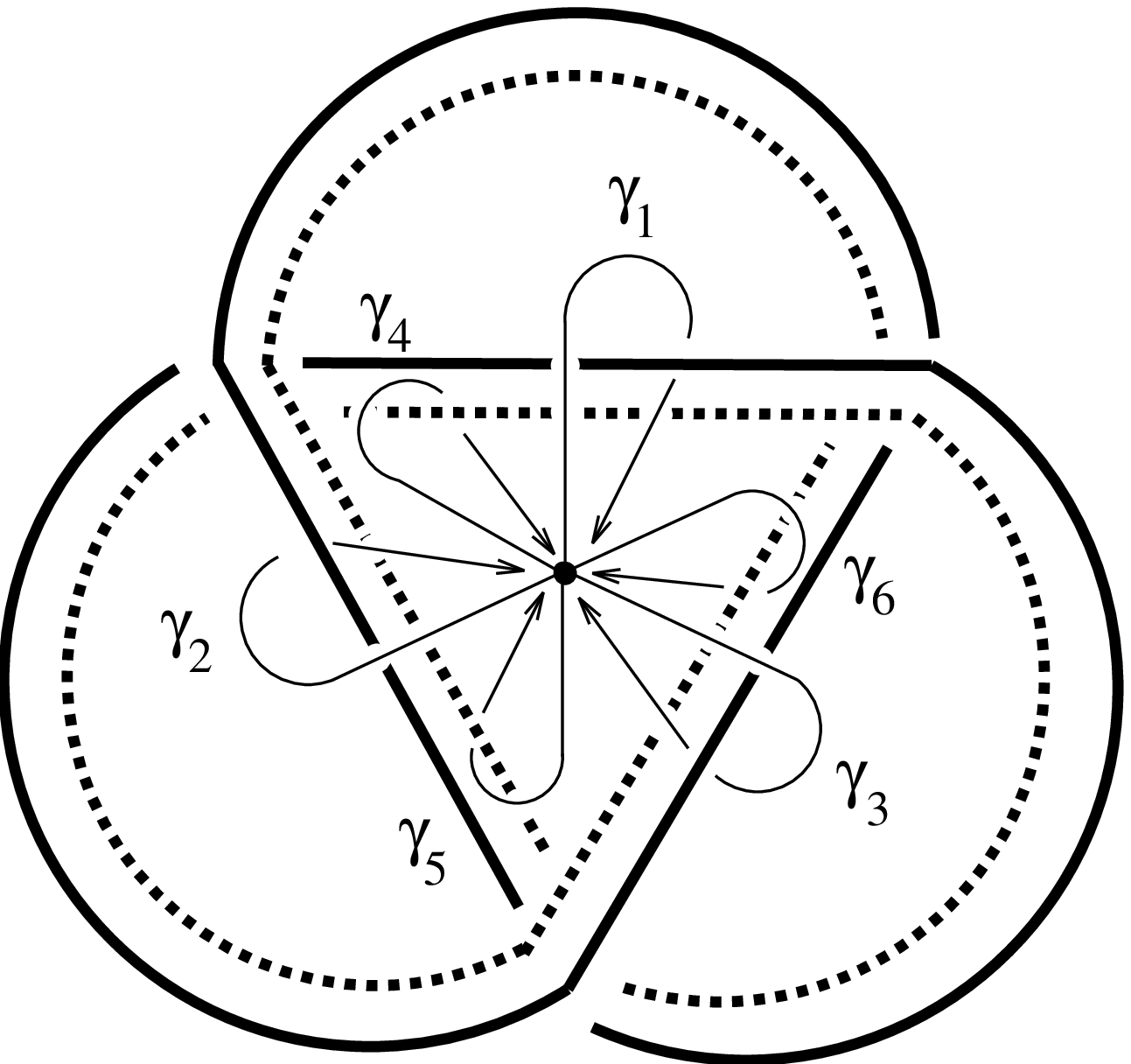}{7cm}

We would now like to compute the monodromies along paths
$\gamma_i$ looping around the knots, shown in Fig.\ 3.  On the one hand,
they can be calculated directly from the one-loop expression for the $SU(3)$
effective action (giving the monodromies along paths $\gamma_i\cdot
\gamma_{i+3}$ whch go around both
knots), and the monodromies of the exact $SU(2)$ solution (which tell how
the $\gamma_i\cdot\gamma_{i+3}$ monodromies must factorize
into the individual monodromies around
each knot separately).  On the other hand, since it was
shown in Ref.\ \AF\ that the curve \curve\ reproduces the field theory
monodromies at weak coupling, we can borrow the results of Ref.\ \KLTYii\
which calculated those monodromies from \curve.  Identifying the
monodromies in terms of the magnetic and electic charges ${\bf n}^{(i)}
= ( h^{(i)} ; q^{(i)} )$ of the massless dyon on the curve encircled by
the path $\gamma_i$, one has:
		\eqn\morechrgs{
	\matrix{{\bf n}^{(1)}&=&(&1,&0;&1,&0&),\cr
	{\bf n}^{(2)}&=&(&0,&1;&-1,&1&),\cr
	{\bf n}^{(3)}&=&(&1,&1;&0,&1&),\cr}\qquad
	\matrix{{\bf n}^{(4)}&=&(&1,&0;&-1,&1&),\cr
	{\bf n}^{(5)}&=&(&0,&1;&0,&-1&),\cr
	{\bf n}^{(6)}&=&(&1,&1;&-1,&0&).\cr}
		}
As there are charges on a single curve which are not mutually local,
for example $\t{\bf n}^{(1)}\cdot {\bf I}\cdot {\bf n}^{(2)} \neq 0$,
we have completed the argument showing that vacua with massless mutually
non-local dyons must occur in the $SU(3)$ (and therefore $SU(N)$)
theories.

\subsec{Degenerations and Monodromies}

We now discuss the way the singular curves on our 3-sphere may cross or
otherwise degenerate as the 3-sphere is deformed, and what monodromies
are allowed at these degenerations.  This discussion is not necessary
for the argument completed above, but it may serve to make some of its
topological aspects more concrete.

3--dimensional slices of 2--dimensional surfaces in a 4--dimensional
space are generically 1--dimensional curves.  As the 3--dimensional
slice is deformed the curves move, and two such curves may touch at a
point.  The stable sequences of such degenerations are shown in
Fig.\ 4.  `Stable' here means that the degeneration can not be
removed by any small deformation of the curves.  The (a) and (b)
degenerations depicted in Fig.\ 4 can be visualized in three dimensions
as a sequence of intersections of a plane with a sphere as the plane
lifts off the sphere, for (a), and as a plane passing through the
saddle point of a saddle (b).  This shows that there is actually no
invariant meaning to the degeneration point in cases (a) and (b).  The
(c) degeneration, on the other hand, is a truly 4--dimensional
phenomenon, being the depiction of the transverse intersection at a
point of two 2--dimensional surfaces.  In this case the degeneration
point is the point of intersection, and has a slice-independent
meaning.

\fig{The three stable degenerations of curves in a 3--dimensional
space.  When viewed as the intersection of 2--surfaces with a sequence
of 3--dimensional slices of a 4--dimensional space, only in case (c)
does the point of intersection have a slice-invariant meaning.}
{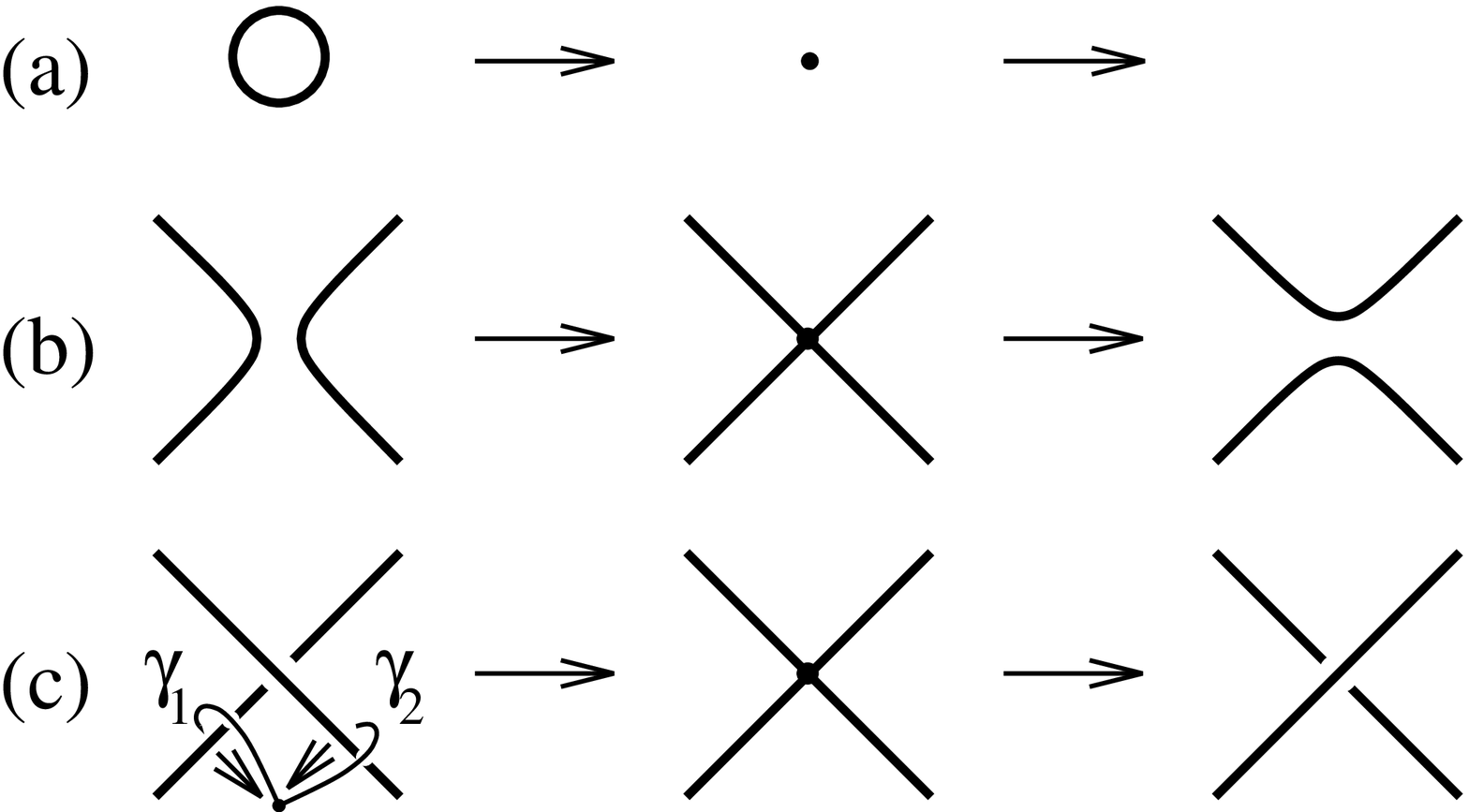}{9cm}

The existence of such degenerations constrains the
possible monodromies around these curves.  For example, around the two
paths $\gamma_1$, $\gamma_2$ marked in Fig.\ 4(c), there may, in principle
be two independent monodromies $\bM_1$ and $\bM_2$.  However, a sequence of
deformations of $\gamma_1$ as we pass through the degeneration and then
back again shows (see Fig.\ 5) that $\gamma_1=\gamma_2\cdot\gamma_1 \cdot
\gamma_2^{-1}$, and thus that $[\bM_1,\bM_2]=0$.  Thus only surfaces with
mutually local massless dyons are allowed to intersect transversely.
Simpler sequences of path deformations show in the (a) and (b)
degeneration cases that there is only {\it one} independent monodromy,
consistent with the fact that these degenerations result from a sequence
of 3--surfaces slicing a single smooth 2--surface.

\fig{Deformation of the $\gamma_1$ path to the path
$\gamma_2\cdot\gamma_1\cdot\gamma_2^{-1}$ in the vicinity of the
crossing depicted in Fig.\ 4(c).}{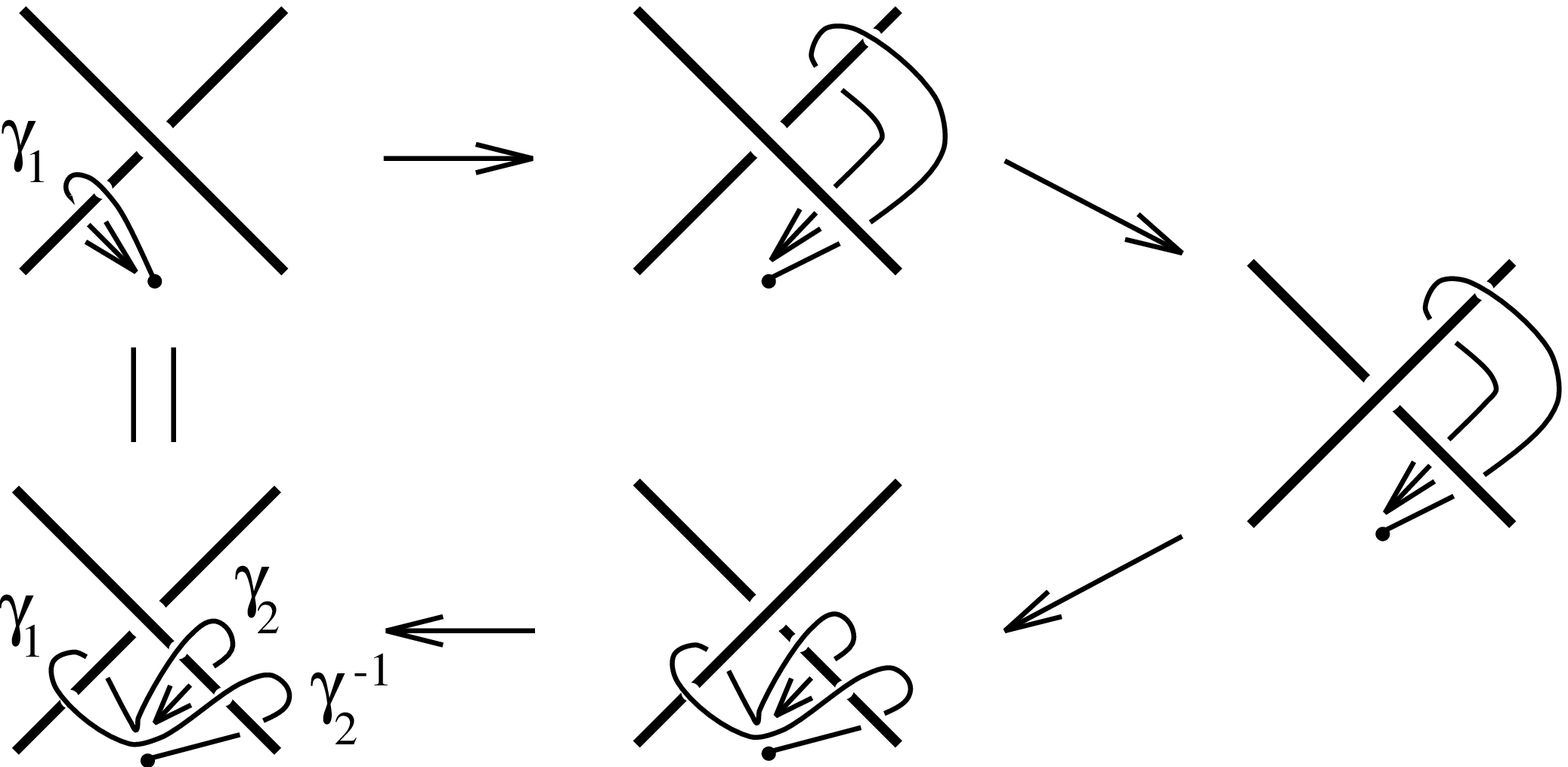}{9cm}

One can now see how the two knotted curves at weak coupling shown in
Fig.\ 3 can unlink as we shrink the 3-sphere to stronger couplings:
the knot represented by a dotted line shrinks until its three loops
cross the other (solid) knot in three places.  These type (c) crossings
are allowed, because \morechrgs\ implies that the relevant monodromies
commute: $[\bM_1,\bM_5]=[\bM_2,\bM_6]=[\bM_3,\bM_4]=0$.  Indeed, these
three crossings correspond to the three $\bZ_2$ vacua with mutually
local massless dyons found from the curve \curve.  The two unlinked
knots can not unknot by any of the stable degenerations of Fig.\ 4,
since we have shown that those degenerations do not support non-commuting
monodromies.  Thus the unknotting must occur via an unstable
degeneration of some sort.  It is not hard to see that of all the
degenerations involving three curves, only the one shown in Fig.\ 6
allows monodromies satisfying the constraints \monrel\ implied by the
topology of the knot.  This degeneration is in fact the one that occurs at
the $\bZ_3$ vacua with mutually non-local massless dyons found above from
the exact solution \curve\ of the $SU(3)$ theory.

\fig{An unstable crossing of three curves at a point in three dimensions.
This can be viewed as successive 3-dimensional slices through the massless
dyon 2--surfaces in the neighborhood of a $\bZ_3$ point in the
$SU(3)$ moduli space.}{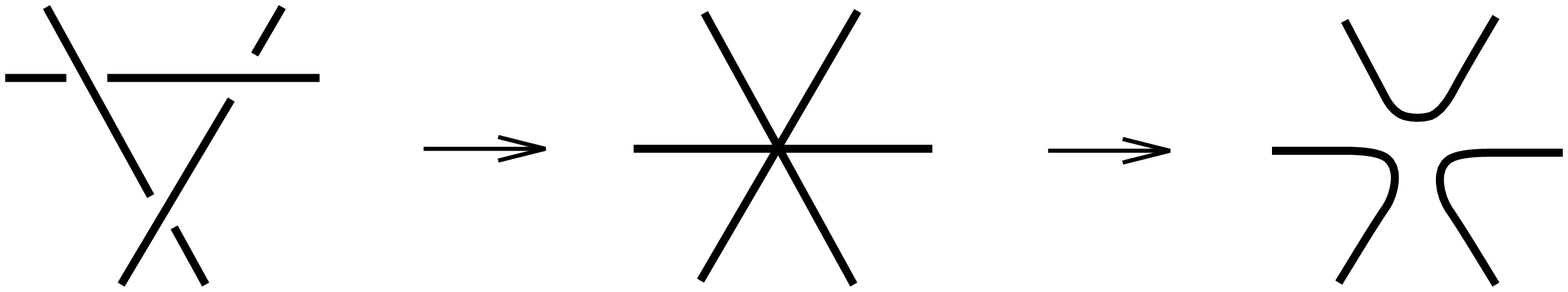}{9cm}

\newsec{Effective Lagrangian near the $\bZ_3$ Vacua}

The low-energy effective Lagrangian of $\CN=2$
$SU(3)$ gauge theory, written in terms of two $U(1)$ gauge multiplets and
the hypermultiplets for the light matter in that region of moduli space,
was given in \effL, \defatau\ and \BPSeff.  The effective couplings
were given by derivatives of an analytic prepotential $\CF$.
One can choose any two symplectically orthogonal charge vectors as the gauge
charges with local couplings, and ordinarily one does this to make the
couplings to the hypermultiplets local.

Near the $\bZ_3$ point, two non-symplectically orthogonal periods
become small: call them $a_1$ and $a_D^1$.  These correspond to the
mutually non-local monopoles which become massless there, since
all masses of light particles are determined by these two ``short''
periods.  To write the matter Lagrangian we need the
non-symplectically orthogonal gauge multiplets $(A_1,W_1)$ and
$(A_D^1,W_D^1)$, containing non-locally related vector potentials.
We will see below that $a_1$ and $a_D^1$
are good coordinates in this region of moduli space.
To write a standard $\CN=2$ Lagrangian (in terms of mutually local
fields) we must choose one of the short periods,
say $A_1$, and one of the ``long'' periods $A_2$ as our variables.
We then have $A_D^1=\p\CF(A_1,A_2)/\p A_1$ by \defatau.


The gauge kinetic term is then determined by $\tau^{ij}$,
the period matrix of the $SU(3)$ quantum curve
		\eqn\surf{\eqalign{
	y^2 &= P(x)^2 - \Lambda^6\cr
	P(x) &= \half \left(x^3 - u x - v\right)
		}}
near one of the $\bZ_3$ points $u=0$, $v=\pm 2\Lambda^3$.  Near the
$\bZ_3$ point with $v=2\Lambda^3$, the branch points satisfy $x^3 -
\delta u~x - 2\Lambda^3 -\delta v = \pm 2\Lambda^3$.  Taking the minus
sign, we see the degenerating branch points approach $x=0$ as
		\eqn\branchtwo{
	x^3 - \delta u~x  -\delta v = 0,
		}
while the plus sign gives three branch points at $x^3=4$.  These six
branch points and a choice of basis of conjugate cycles is shown in
Fig.\ 7.

\fig{The distribution of the six branch points in the $x$-plane
near a $\bZ_3$ point.  Three points are of order $\delta u^{1/2}$ or
$\delta v^{1/3}$ from the origin, while the other three are close to
$x=e^{2\pi i k/3} 2^{2/3} \Lambda$.  The dotted lines are a choice of
branch cuts, and a basis of conjugate cycles is shown, with
the solid and dashed lines on the first and second sheets.}
{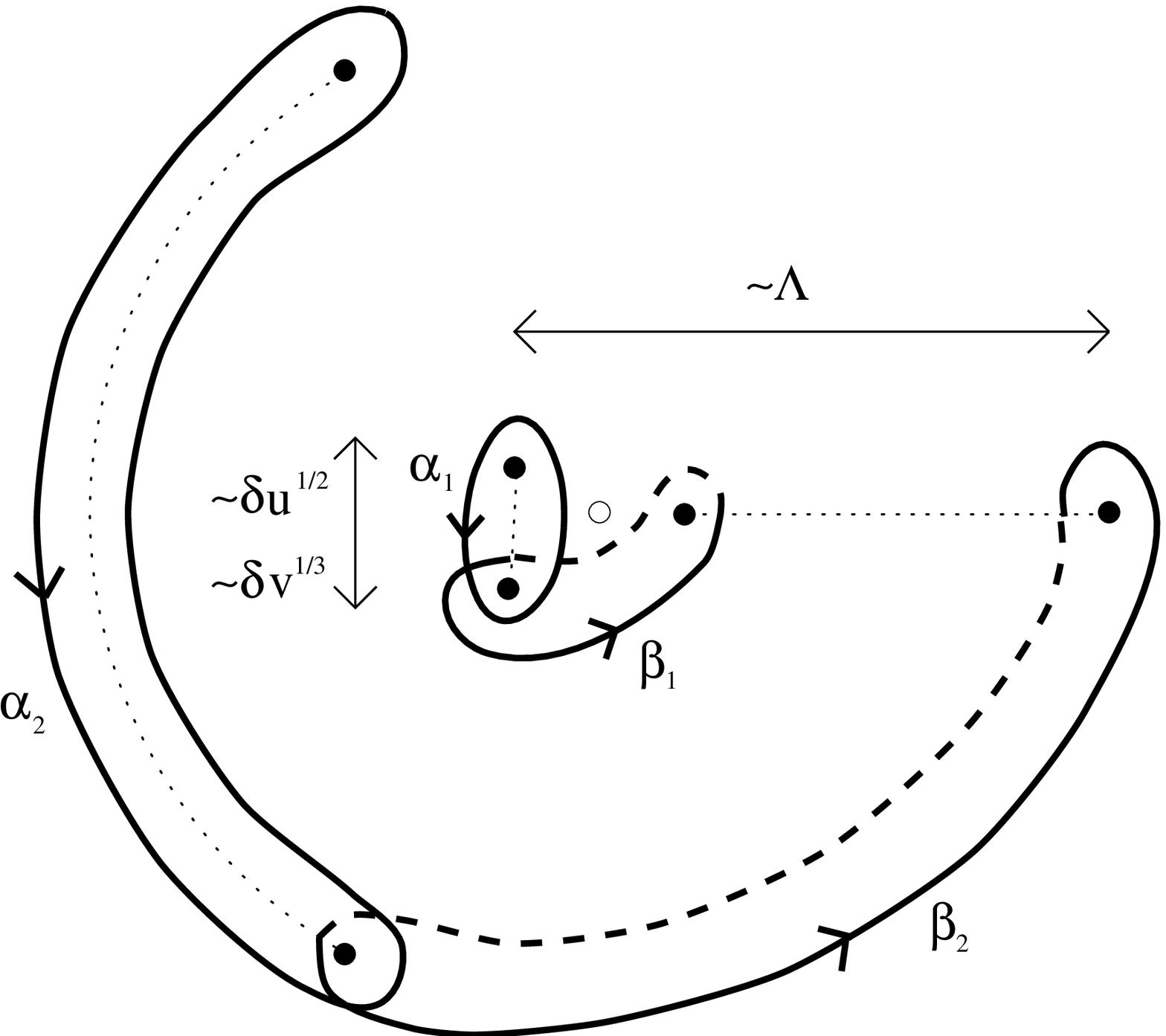}{8cm}

The short periods $(a_1,a_D^1)$ and the long periods
$(a_2,a_D^2)$ are given by
		\eqn\shrtper{
	a_i = \oint_{\alpha_i} \lambda ,\qquad
	a_D^i = \oint_{\beta_i} \lambda ,
		}
where the one-form $\lambda$ is given by \meroform.
Near the $\bZ_3$ point as the short periods vanish,
the corresponding light states are in this basis an
electron with mass $|a_1|$, a monopole with mass $|a_D^1|$,
and a dyon of mass $|a_1+a_D^1|$.  These are all charged only
with respect to the first $U(1)$ factor.

Close to the $\bZ_3$ point, the
two pairs of conjugate cycles $(\alpha_1,\beta_1)$ and
$(\alpha_2,\beta_2)$ are separated by a ``neck'' whose conformal
parameter is becoming large, as illustrated in Fig.\ 8.
This degeneration of a genus two Riemann surface was studied in
\lebo\ and several of the following results are derived there.
This is a
familiar limit in string theory and would correspond there to two
`one-loop tadpole' amplitudes connected by a zero momentum
propagator.  In this basis the period matrix $\tau^{ij}$ splits
at the degeneration point as
		\eqn\permat{
	\tau^{ij} = \pmatrix{\tau^{11} & 0\cr 0&\tau^{22}\cr}
		}
where $\tau^{11}$ is the modulus of the ``small'' torus and
$\tau^{22}$ is the modulus of the ``large'' torus.

\fig{The degeneration of the genus 2 Riemann surface corresponding to the
distribution of branch points shown in Fig.\ 7 as $\delta u$, $\delta v
\rightarrow 0$.}{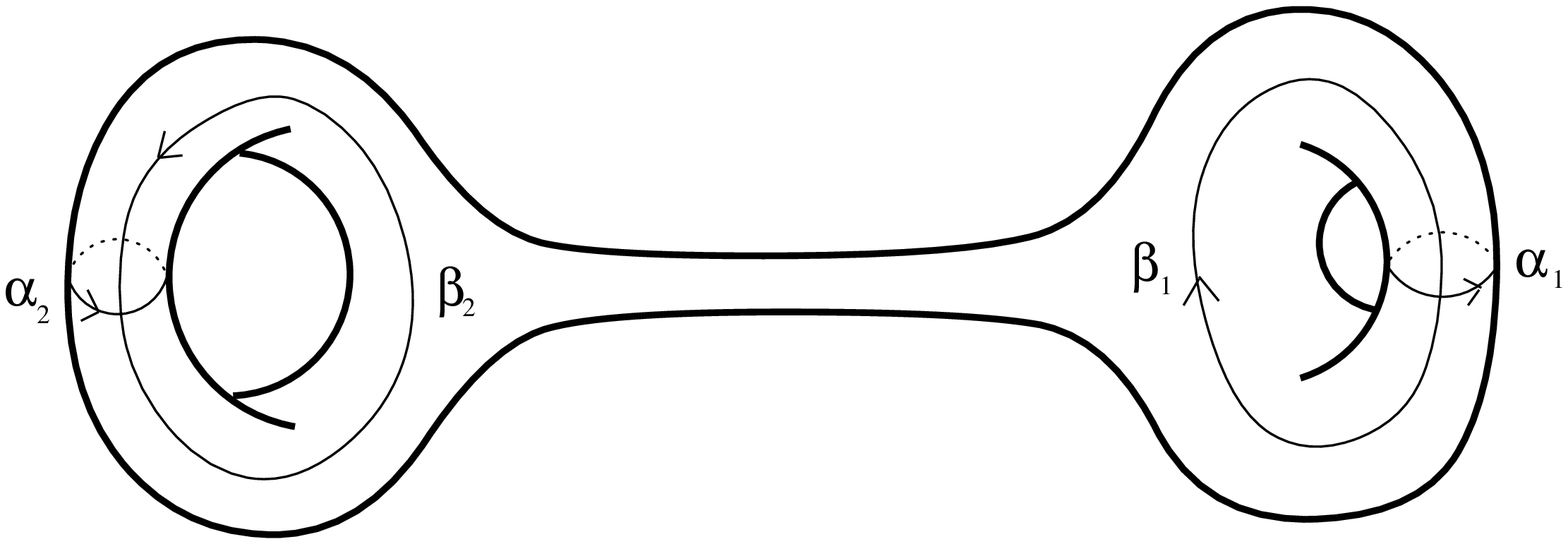}{10cm}

The `large' torus is obtained by simply identifying the three
degenerating points and taking a double cover of the resulting surface
with four punctures.  Thus, near the $\bZ_3$ point it will be described
by the curve $w^2=x(x^3-4\Lambda^3)$ plus order $\delta u/\Lambda^2$
and and $\delta v/\Lambda^3$ corrections.  Because of the $\bZ_3$
symmetry of this curve as $\delta u$, $\delta v\rightarrow 0$, near the
$\bZ_3$ point the modulus of the large torus is
		\eqn\largetorus{
	\tau^{22}=e^{2\pi i/6}
	+\CO\left(\delta u\over\Lambda^2\right)
        +\CO\left(\delta v\over\Lambda^3\right).
		}

The ``small'' torus can be obtained similarly after a conformal
transformation; it is described by the curve
		\eqn\smalltorus{
	w^2=x^3-\delta u x-\delta v
		}
plus order $\delta u/\Lambda^2$ and $\delta v/\Lambda^3$ corrections
near the $\bZ_3$ point.  In the limit its modulus is
		\eqn\smalltau{
	\tau^{11}=\tau(\rho)
	+\CO\left(\delta u\over\Lambda^2\right)
	+\CO\left(\delta v\over\Lambda^3\right).
		}
The function $\tau$ depends only on  $\rho^3 \equiv \delta u^3/\delta v^2$
(as can be seen by rescaling $x$),
or equivalently, the angles of the triangle formed by the degenerating
points.

The monodromy around a path encircling, say, the $a_1=0$ curve near
the $\bZ_3$ point take $a_1\rightarrow e^{2\pi i} a_1$, $a_D^1 \rightarrow
a_D^1 + a_1$, and leaves the long periods unchanged.  Since the action
of this monodromy on $\tau^{ij}$ is to shift only $\tau^{11}\rightarrow
\tau^{11}+1$ and leave
$\tau^{12}$ and $\tau^{22}$ unchanged, we see that the off-diagonal
terms of the period matrix are analytic in the short periods.
Since the period matrix splits, the off-diagonal terms must also
vanish at the $\bZ_3$ point, so we have
		\eqn\split{
	\tau^{12} = \tau^{21} = \CO\left( a_1\over\Lambda\right)
	+ \CO\left( a_D^1\over\Lambda\right).
		}
Physically, the splitting of the period matrix will mean that the two
$U(1)$ gauge factors decouple.  Integrals of $\lambda$ over the long
periods will all be $O(\Lambda)$ and thus all particles with charge in
the second $U(1)$ will have masses of $O(\Lambda)$.  On the other hand,
particles with charge only in the first $U(1)$ will have masses going
to zero in the limit, producing a separation in scales between the two
sectors.  Let us keep this in mind, but defer most of the physical
interpretation to the next subsection.

We thus need to compute $\tau$, $a_1$, $a_D^1$ and $a_2$ as
functions of $\delta u$ and $\delta v$, and then re-express $\tau$
in terms of $a_2$ and one of the short periods.  The long period $a_2$
depends analytically on these parameters:  $a_2 \sim \Lambda + \delta
u/\Lambda + \delta v/\Lambda^2 + \ldots$ When the separation of scales
is large, this dependence will be weak and we can trade $\delta u$ for
$a_2$.

Useful parameters for the small torus are $\delta v=2 \epsilon^3$
and $\delta u = 3\epsilon^2 \rho$.  The overall mass scale is
$\epsilon$, while the dimensionless $\rho$ determines the modulus of
the small torus in the limit: rescaling $x=\epsilon z$, the defining
equation of the small torus near the $\bZ_3$ point \smalltorus\ becomes
		\eqn\smalltor{
	w^2=z^3-3\rho z-2,
		}
depends only on $\rho$ and not on $\epsilon$.  The natural
dimensionless parameter is actually $\rho^3$ since the moduli space of
the curve \smalltor\ has a $\bZ_3$ symmetry $\rho\rightarrow e^{2\pi
i/3}\rho$.  The interesting points in its moduli space are as follows:
the degeneration of the small torus at $\rho^3=1$,
the point $\rho^3=0$ where the small
torus has a $\bZ_3$ symmetry, and the point $\rho^3=\infty$ where it
has a $\bZ_2$ symmetry (rescale $z\rightarrow \rho^{1/2} z$ to see this
behavior).
The $\rho^3$ plane is an $SL(2,\bC)$ transformation of the
$j$ (fundamental modular invariant) plane, $j=27\cdot64\rho^3/(\rho^3-1)$.

We wish to compute the periods of the form $\lambda\propto (x/y) dP$ on
the small torus.  It degenerates to
		\eqn\deglam{
	\lambda \propto
	{\epsilon^{5/2}\over \Lambda^{3/2}} {z (z^2-\rho)\over w}~dz
	= {\epsilon^{5/2}\over \Lambda^{3/2}} w~dz + d(\ldots),
		}
on \smalltor.  We will start by examining the periods near $\rho^3=1$,
$\infty$, and $0$, in turn.

If $\rho$ is near a
degeneration of the small torus, $\rho^3=1+\delta\rho$, the three branch
points are at $z=-1\pm\CO(\delta\rho^{1/2})$ and $z=2+\CO(\delta\rho)$.
Call the periods $a_s$ and $a_l$ with $|a_s|\le|a_l|$.
Then, from \deglam, one finds
		\eqn\pshort{
	a_s \propto (\delta\rho) {\epsilon^{5/2}\over\Lambda^{3/2}},\qquad
	a_l \propto {\epsilon^{5/2}\over\Lambda^{3/2}}.
		}
Thus we find that one of the three light hypermultiplets has a mass
$m_s/\Lambda \sim \delta\rho (\epsilon/\Lambda)^{5/2}$, while the
other two have masses $m_l \sim (\epsilon/\Lambda)^{5/2}$.  Which
of the three particles is lightest depends on which
root of $\rho^3=1$ we are expanding about:  the $\bZ_3$ symmetry
of the small torus \smalltor\ acts on the periods as $a_1\rightarrow
a_D^1\rightarrow -a_1-a_D^1$.

The modulus $\tau(\rho)$ of the small torus in this limit can be
determined from the monodromies to be
		\eqn\lowbeta{
	\tau = {1\over 2\pi i}\log {a_s\over a_l} + \ldots
		}
since, as mentioned above, along a path $a_1 \rightarrow e^{2\pi i}
a_1$ encirlcing the $a_1=0$ curve near the $\bZ_3$ point $\tau
\rightarrow \tau + 1$.  Taking $a_1=a_s$, this gives \lowbeta, since
$\tau$ depends only on $\rho$.  Alternatively, one can calculate
\lowbeta\ directly from the definition of the modulus in terms
of the periods of the holomorphic one-form $dz/w$ on the
small torus: $\tau = \oint_{\beta_1}{dz\over w}/\oint_{\alpha_1}
{dz\over w}$.

The most important feature of the result is that the final low energy
coupling is independent of $\epsilon/\Lambda$, which scaled out of the
modulus of the small torus.  This is true for all $\rho$.  We save a
detailed discussion of the physical interpretation for the next
section, but the simplest interpretation of this result is that it
comes from integrating the beta function for a single charge one
hypermultiplet, turning on at the scale $a_l$, the mass of the two
heavier particles, and turning off at the scale $a_s$, the mass of the
light particle.

The limit in which $\rho\rightarrow\infty$ is also interesting.  This
limit sends two of the branch points of the small torus to infinity as
$z\sim \pm\sqrt{\rho}$, while the third stays at the origin.
Therefore, this is is not a degeneration, because it can be undone by
the rescaling $z\rightarrow \rho^{1/2} z$.  The limit is the torus with
modulus $\tau=i$ and $\bZ_2$ symmetry.  The periods are $a_\pm=C
(\epsilon^2\rho)^{5/4}/\Lambda^{3/2} i^{\pm 1/2}$ and the three
particles have mass $|a|$, $|a|$ and $\sqrt{2}|a|$.

Finally, taking $\rho=0$ produces a small torus with $\bZ_3$ symmetry,
and modulus $\tau = e^{2\pi i/3}$.  The periods are $a_1=e^{2\pi
i/3}a_D^1=C'\epsilon(\epsilon/\Lambda)^{3/2}$ and all three particle
masses are equal.

In summary, we have developed a picture of the vicinity of the $\bZ_3$
point, shown in Fig.\ 9.  The limit $\epsilon/\Lambda \rightarrow 0$
takes us to the $\bZ_3$ point, and varying $\rho$ changes the direction
from which we approach it.  The masses all vary with $\epsilon$ as
$m \sim (|a_1|, |a_D^1|, |a_1+a_D^1|) \sim \epsilon^{5/2}/\Lambda^{3/2}$.
The $\Im a^1_D/a_1=0$ curve will be determined below.

\fig{Map of the vicinity of the $\bZ_3$ point, in the coordinates $\rho$
and $\epsilon/\Lambda$.  The latter is the (complex) dimension out of the
page.}
{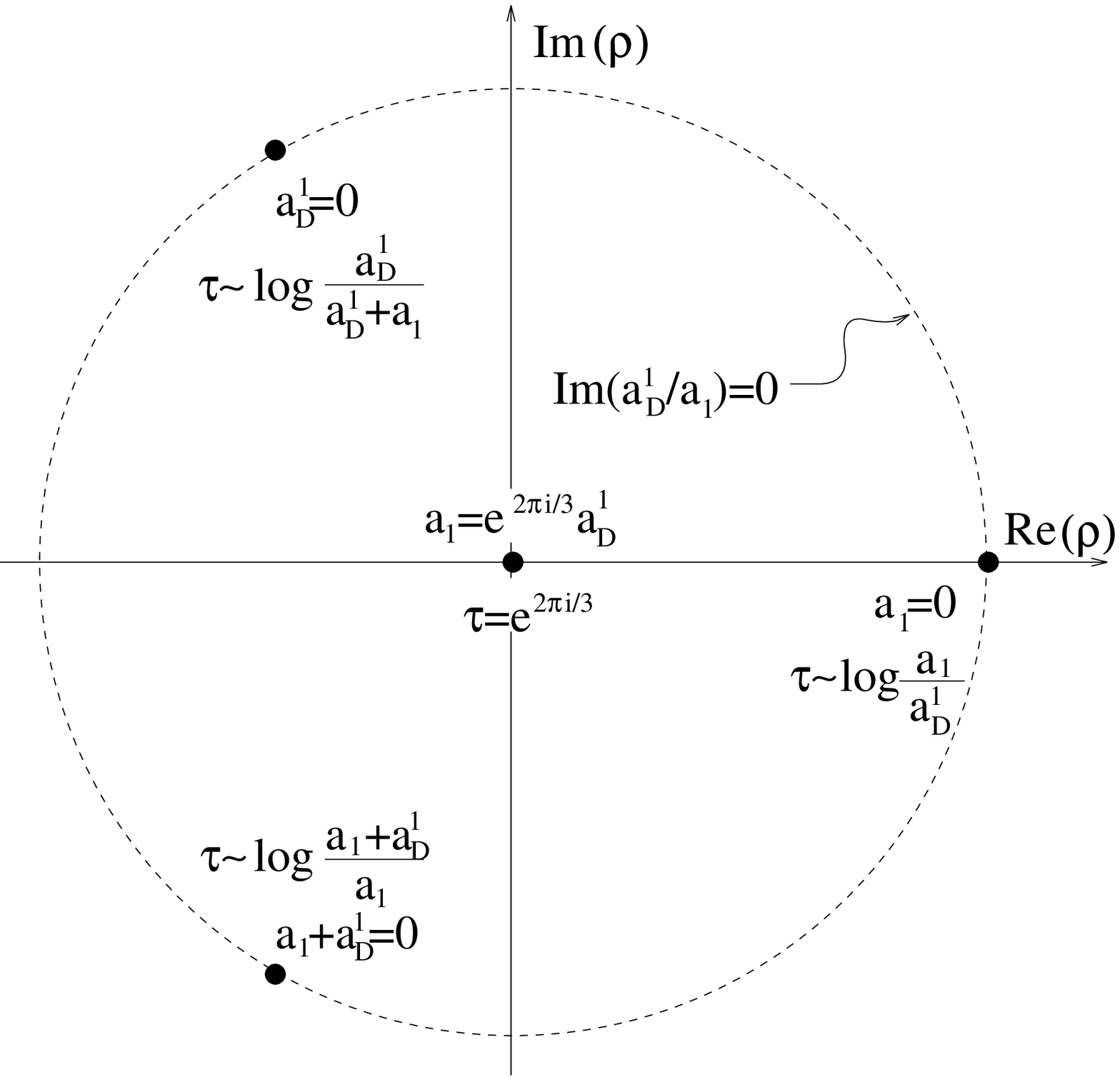}{10cm}

\subsec{Elliptic function representation}

To describe the whole moduli space, a representation in terms of
elliptic functions is useful \Ell.

The parameters describing the small torus, $\delta u$ and $\delta v$ of
\branchtwo, are almost the standard parameters of elliptic function
theory:  $g_2 = 4\delta u$ and $g_3=4\delta v$.  The small torus
becomes
		\eqn\branchtwob{
	w^2 = 4 x^3 - g_2~x - g_3.
		}
Another pair of parameters we can use are the two periods of the holomorphic
form $dx/w$, to be called $\omega$ and $\omega_D$.~\footnote*{
In many references, the periods are $2\omega$ and $2\omega_D$.}
Their ratio is $\tau=\omega_D/\omega$, while their overall scale is
related to $\epsilon$ as $\epsilon^{-1/2}$.  The short periods $a_1$ and
$a_D^1$ are the
periods of the form $\tilde\lambda =  w~dx$.  The $\bZ_3$ symmetry
is $(\omega,\omega_D)\rightarrow (\omega_D,-\omega-\omega_D)$ and acts
the same way on $(a_1,a_D^1)$.
The relation between $(g_2,g_3)$ and $(\omega,\omega_D)$ is standard:
		\eqn\geisen{\eqalign{
	{g_2\over 60} &= \sum_{m,n}{}' {1\over (m\omega+n\omega_D)^4}
	 = {\pi^4\over 45\omega^4}\left(1+240q+\ldots\right)\cr
	{g_3\over 140} &= \sum_{m,n}{}' {1\over (m\omega+n\omega_D)^6}
	= {2\pi^6\over 35\cdot27\omega^6}\left(1-504q+\ldots\right).
		}}
Using properties of these elliptic functions it is shown in the
Appendix that
		\eqn\nicerel{
	a_1 = -{\pi i\over 5}{\p g_2\over\p\omega_D},\qquad
	a_D^1 = {\pi i\over 5}{\p g_2\over\p\omega},\qquad
	\omega_D a_1 - \omega a_D^1 = {4\pi i\over 5} g_2.
		}

The degeneration $\tau\rightarrow i\infty$ takes
		\eqn\deginf{
	a_1 \sim {128\pi^6\over\omega^5}e^{2\pi i\tau} \ \rightarrow 0,
	\qquad\qquad a_D^1 \rightarrow  - i{16\pi^5\over 15 \omega^5},
		}
reproducing the result \lowbeta.

We now derive the prepotential $\CF(a_1,a_2)$.  A long period $a_2$
depends only weakly on the degeneration parameters.  It can be
calculated by expanding $\lambda$ in $\delta u$ and $\delta v$, and it
will depend analytically on them:
		\eqn\longper{
	a_2 = \hat{a_2} + C {\delta u\over \Lambda}
	+ \CO\left({\delta v\over \Lambda^2}\right)
	+ \CO\left({\delta u^2\over \Lambda^3}\right),
		}
where $\hat{a_2}\sim\Lambda$ is the value of the long period at the
$\bZ_3$ point, and $C$ is some calculable constant.  If we take $\delta
u\sim\epsilon^2$ and $\delta v\sim\epsilon^3$, the $\delta v$
dependence is subleading, so we can trade $a_2$ for $g_2=4\delta u$ by
writing $g_2 = (4/C)\Lambda(a_2-\hat{a_2})$.  A small variation of $a_2$
translates into a large variation of $g_2$ and thus of $\rho$ and
$\tau$: $\p\tau/\p a_2 \sim \Lambda/\epsilon^2$.

The results \nicerel\ implicitly define the prepotential $\CF(a_1,g_2)$
and the functions $\tau(a_1,g_2)$ and $a_D^1(a_1,g_2)$ appearing in the
effective Lagrangian.  A more explicit expression would be too
complicated to be very illuminating, and it is better to think in terms
of the picture in Fig.\ 9.  The limiting behavior of $\CF$ as
$a_1\rightarrow 0$ is simple:  to reproduce $\p\CF/\p a_1=a_D^1$,
		\eqn\cflim{
	\CF(a_1,a_2) \rightarrow
	-i{(3\cdot 64)^{1/4}\over 5}\ a_1 g_2^{5/4}
	\propto a_1 (a_2-\hat{a_2})^{5/4}.
		}
The non-trivial exponent is associated with the scaling $m\sim
\epsilon^{5/2}/\Lambda^{3/2}$ derived earlier.  (Note that this
non-analyticity is not directly reflected in the monodromy---the
non-trivial monodromy visible in this limit is $a_1\rightarrow e^{2\pi
i} a_1$, while in the limit $g_2\rightarrow 0$ this asymptotic form is
not valid.)

\subsec{Stability of BPS states}

As discussed in \SWi, BPS states become marginally stable on lines $\Im
a_D^1/a_1=0$.  We now prove a result needed for section 5:  for small
$\epsilon$, this line is a simple closed curve in the $\rho$ plane,
passing through each cusp, separating $|\rho|>>1$ from $|\rho|<<1$, and
$a_D^1/a_1$ can take any real value.


$a_D^1/a_1$ transforms under $SL(2,\bZ)$ in the same way as $\tau$.
The $\rho$ plane can be mapped to the region $0\le\Re \tau\le 1$,
$|\tau-\half|>\half$.  Its image in the $a_D^1/a_1$ plane is the region
$R'$ shown in Fig.\ 10, and the dashed line in the region $R$ is the
preimage of $\Im a_D^1/a_1=0$.

\fig{The complex $\tau$ and $a_D^1/a_1$ planes.  The modular domains
mapped onto the $\rho$ plane are shown.  The dashed curve is the image
of the $\Im a_D^1/a_1 =0$ curve in the $\tau$ plane.}
{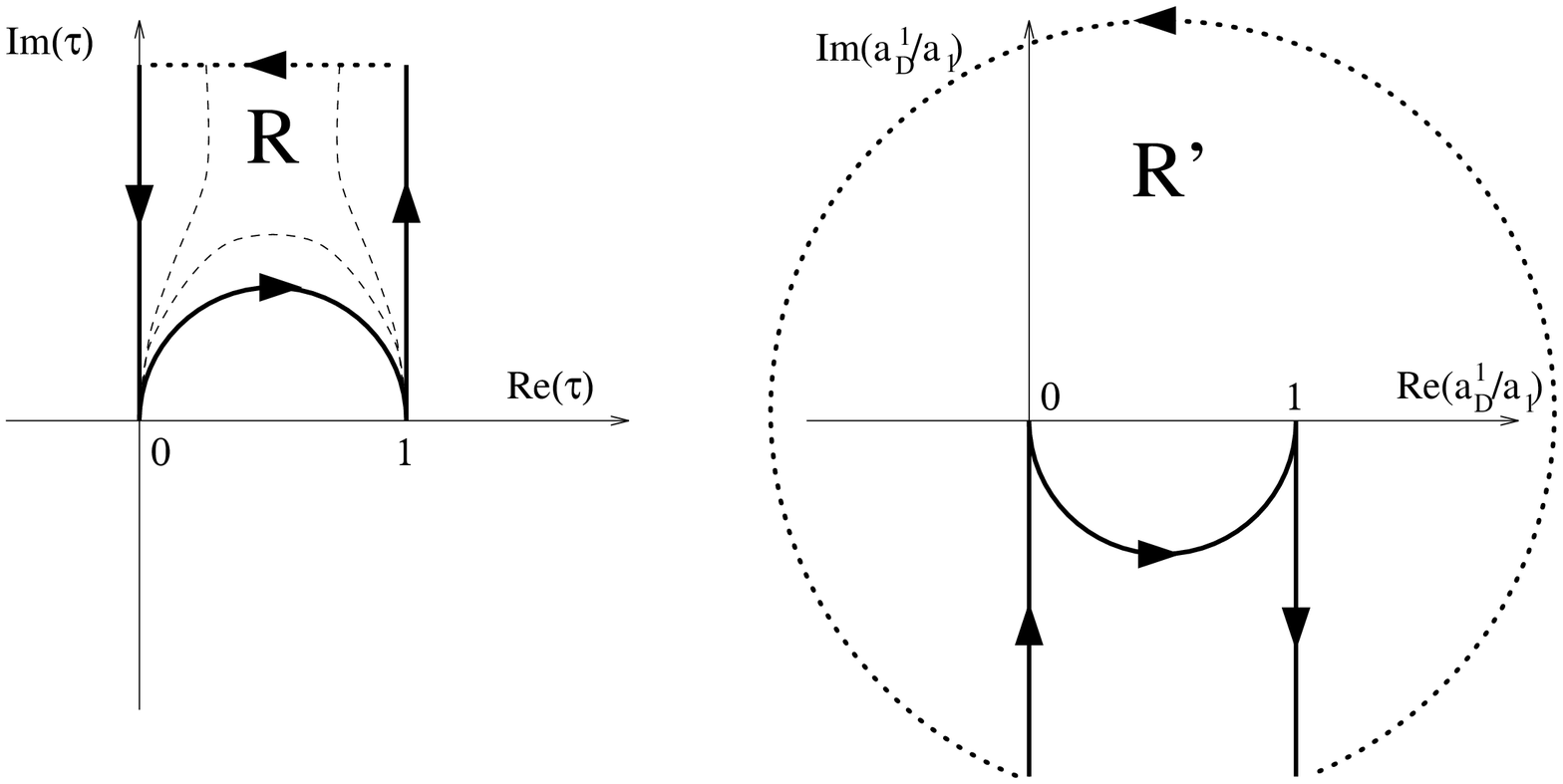}{11cm}

This can be seen as follows.  First, for large $\Im\tau$, $a_D^1/a_1 =
\tau + (1/120\pi)e^{2\pi \Im\tau}e^{i \pi(3/2-\Re \tau)}$ and a line
$0\le \Re\tau\le 1$ with $\Im\tau$ fixed sweeps out a large circle in
the $a_D^1/a_1$ plane.  The images of $\tau=0$ and $\tau=1$ are known,
and from the expression for $a_D^1/a_1$ it is clear that the outer
lines run parallel to the imaginary axis.

We then observe that if two points in the $\tau$ plane are related by
$SL(2,\bZ)$ as $\tau_1=g(\tau_2)$, their images in the $a_D^1/a_1$
plane will have the same relation.  $\tau=i$ is a fixed point of
$z\rightarrow -1/z$, so its image is $a_D^1/a_1=-i$.  By applying
$z\rightarrow -1/(z-1)$ we get $\tau=\half(1+i)$, so its image is
$a_D^1/a_1=\half(1-i)$.  Finally, the quarter arcs $\tau=0$ to
$\tau=\half(1+i)$ and $\tau=1$ back to $\tau=\half(1+i)$ and their
images are exchanged by the $SL(2,\bZ)$ transformation $z\rightarrow
{z-1\over 2z-1}$, and related by $\Re\tau\rightarrow 1-\Re\tau$, which
determines them.

\subsec{Breaking to $\CN=1$}

We now consider what happens to the $\CN=2$ $SU(3)$ gauge theory when
we add terms to the microscopic superpotential explicitly breaking
$\CN=2$ to $\CN=1$ supersymmetry.  We will not use these results in an
essential way, but it is interesting to see that the resulting $\CN=1$
theory will have ground states near the $\bZ_3$ point, and that the
pure $\Tr\phi^3$ superpotential makes the $\bZ_3$ point itself a ground state.

Near a point in moduli space where two or fewer mutually local dyon
hypermultiplets are massless we can, by a duality transformation on the
low energy gauge fields, choose them to be separately electrically
charged with respect to the two $U(1)$'s.  Denote by $A_i$, $i=1,2$ the
$\CN=1$ chiral superfield parts of the two $\CN=2$ $U(1)$ gauge
multiplets, and by $a_i$ the vevs of the lowest components of $A_i$.
Then locally the $a_i$ are coordinates on the moduli space which vanish
where the dyon charged with respect to their $U(1)$ is massless.  In
terms of the $\CN=1$ chiral superfields $M_i$, $\tilde M_i$ which form
the dyon hypermultiplets, the $\CN=2$ superpotential becomes
              \eqn\dyonw{
      \CW_0 = \sqrt2 \sum _{i=1}^\ell A_i M_i \tilde M_i.
              }

In the microscopic superpotential we can add the two renormalizable terms
$(\mu/2) \tr \Phi^2 + (\nu/3) \tr \Phi^3$, which break $\CN=2$ to $\CN=1$.
For small $\mu$ and $\nu$, the superpotential in the low-energy theory is
then
              \eqn\lew{
      \CW = \CW_0 + \mu U + \nu V.
		}
where $U$, $V$ are the superfields corresponding to $\tr \Phi^2$ and
$\tr \Phi^3$ in the low energy theory; their first components have the
expectation values $\vev{u}={1\over2}\vev{{\rm tr}\phi^2}$ and $\vev{v}
={1\over3}\vev{{\rm tr}\phi^3}$.  Using the non-renormalization
theorem of \Snr,
an argument like one in \SWi\ shows that \lew\ is the
exact low-energy superpotential.  Since $u$ and $v$ are good global
coordinates on moduli space, it is useful to take them as our basic
chiral fields and consider $A_i(U,V)$ as functions of them.

The vanishing of the $D$-terms imply $|m_i|=|\tilde m_i|$, while setting
$d\CW=0$ gives the vacuum equations
              \eqn\Fterma{\eqalign{
      -{\mu\over\sqrt 2} &= {\p a_1\over\p u} m_1 \tilde m_1
      + {\p a_2\over\p u} m_2 \tilde m_2,\cr
      -{\nu\over\sqrt 2} &= {\p a_1\over\p v} m_1 \tilde m_1
      + {\p a_2\over\p v} m_2 \tilde m_2,\cr
              }}
and
              \eqn\Ftermb{\eqalign{
      a_1 m_1 = a_1 \tilde m_1 &=0,\cr
      a_2 m_2 = a_2 \tilde m_2 &=0.\cr
              }}
Here we have denoted by lower-case letters the vevs of the first
components of the corresponding upper-case superfields.

At a point in moduli space where no dyons are massless,
both $a_i$ are non-zero,
so by \Ftermb\ $m_i=\tilde m_i=0$.  Then \Fterma\ has a solution only if
$\mu=\nu=0$.  Thus we learn that the generic $\CN=2$ vacuum is lifted
by the superpotential.

Now consider the $\bZ_2$ points in moduli space where two mutually local
dyons are massless.  At these points $a_1=a_2=0$, so
the $m_i$ are unconstrained by \Ftermb.  For any $\mu$, $\nu$,
\Fterma\ can be solved by adjusting $m_i \tilde m_i$ appropriately, since
the $a_i$ are non-degenerate coordinates at these points.
Up to gauge transformations this is a single solution, describing a
vacuum with a magnetic Higgs mechanism in both $U(1)$ factors.
These $\CN=1$ vacua persist for all values of the bare couplings
$\mu$ and $\nu$.

The interesting case is at a point with just one massless dyon,
say $M_1$, $\tilde M_1$.  In terms of local coordinates on the $\CN=2$
moduli space, this occurs along the one complex dimensional curve
$a_1=0$, but $a_2 \neq 0$.  In terms of the global coordinates $u$,
$v$, this curve is one of the single dyon singularities
$\Delta(Q_\pm)=0$ discussed in Section 2.  A simplified picture of
these curves is shown in Fig.\ 11.  Eq.\ \Ftermb\ implies $m_2=0$, and
so \Fterma\ has a solution if
              \eqn\odcond{
      {\mu\over\nu} = {(\p a_1/\p u)\over(\p a_1/\p v)},
              }
since $m_1$ can be adjusted freely.  It is easy to see qualitatively
where this condition has solutions.  The components in the $u,v$
coordinate system of the normal vector to the massless dyon curve
$a_1=0$ are $(\p a_1/\p u, \p a_1/\p v)$.  If $\nu=0$, so that there is
only a ${\rm tr}\Phi^2$ breaking term, then a solution exists only at a
point where $\p a_1/\p v=0$.  But, from Fig.\ 11 it is
apparent\footnote*{Actually, Fig.\ 11 suppresses the fact that $u$, $v$,
and $\mu/\nu$ are really complex; however, it is not hard to see that
the complex structure ensures that the conclusions of this
two-real-dimensional reasoning are correct.} that the $v$-component of
the normal vector to any of the single-dyon curves vanishes only at
infinity, and thus there are no new $\CN=1$ vacua (in addition to the
$\bZ_2$ points, which always remain vacua).  Now, if we turn on a small
${\rm tr}\Phi^3$ term, $|\mu/\nu|\gg \Lambda$, then the solutions of
\odcond\ will move in from infinity along the single-dyon curves.  This
is shown in Fig.\ 11:  decreasing $|\mu/\nu|$ corresponds to moving the
dashed line towards $u=0$; its intersections with the single-dyon
curves are the new $\CN=1$ vacua.  Under the discrete global $\bZ_6$
$R$--symmetry, $\mu/\nu \rightarrow e^{i\pi/3}\mu/\nu$ when $u
\rightarrow e^{2\pi i/3} u$ and $v \rightarrow -v$.  Thus, changing the
sign of $\mu/\nu$ will pick out two different $\CN=1$ vacua related to
the first two by $v\rightarrow -v$ with their $u$ coordinates the
same.

\fig{A 2-dimensional section of the $SU(3)$ moduli space, showing the
two massless-dyon curves.  The two cusps are the $\bZ_3$ points, and
their intersection is one of the $\bZ_2$ points.  The dashed
$u=$constant line is determined by the ratio of the $\CN=2$--breaking
parameters $\mu/\nu$.  The open circles denote points which remain
$\CN=1$ vacua upon breaking with those parameters.}
{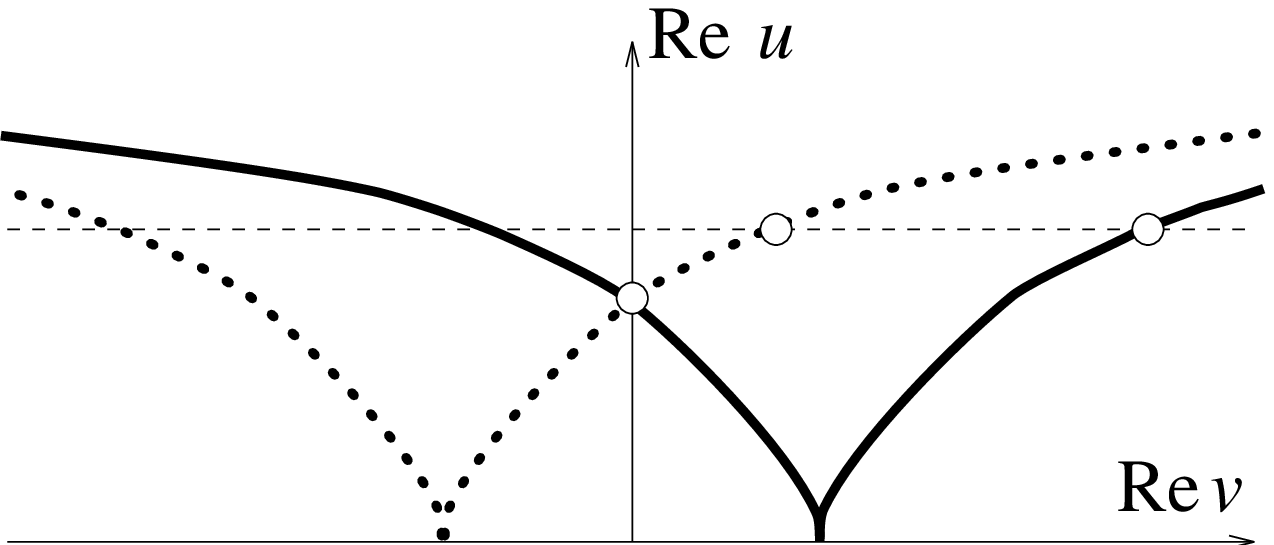}{8cm}

Thus, for generic $\mu/\nu$ we find a total of five $\CN=1$ vacua
(three of which are shown in Fig.\ 11).  Three are the completely
Higgsed $\bZ_2$ points already discussed, while the other two are vacua
in which one $U(1)$ factor is magneticlly Higgsed, while the other
remains unHiggsed since there are no light particles charged with
respect to it.  We will refer to these latter two vacua as the
``half-Higgsed'' vacua.

There are eight special values of $\mu/\nu$ at which this generic
situation no longer holds.  The first is $\mu/\nu=\infty$, discussed
above, where only the three $\bZ_2$ vacua are not lifted.  The next six
are $\mu/\nu=e^{i\pi k/3}\Lambda$,  $k=1,\ldots,6$, where one of the
half-Higgsed vacua coalesces with the $\bZ_2$ vacuum at $v=0$,
$u = e^{2i\pi k/3}3\Lambda^2$.  It is easy to see from the explicit
computations of \DS\ near the $\bZ_2$ points that at, say, the $v=0$,
$u=3\Lambda^2$ vacuum ${\p a_1\over \p u}/{\p a_1\over \p v} =
- {\p a_2\over \p u}/{\p a_2\over \p v} = \Lambda$.  When
$\mu/\nu= \pm \Lambda$ the $\CN=1$ vacuum equations \Fterma\ have
a solution with one of the $m_i=0$. This is a half-Higgsed vacuum with
a massless dyon in the unHiggsed $U(1)$.  Thus at these points in parameter
space there are four $\CN=1$ vacua: two Higgsed and two half-Higgsed with
one of the latter having an extra massless charged particle.

The last special value in parameter space is $\mu=0$, which is a pure
${\rm tr}\phi^3$ breaking term.  From Fig.\ 11 it is apparent that as
$\mu/\nu\rightarrow 0$, the two half-Higgsed vacua approach the $\bZ_3$
points.  Near the $\bZ_3$ points, the condition \odcond\ for an $\CN=1$
vacuum can be evaluated explicitly.  The $a_1=0$ curve corresponds
to the $\rho=1$ point in the $\rho$--$\epsilon$ parametrization of
the moduli space (recall Fig.\ 9).  The behavior of the degenerating
period $a_1=a_s$ near $\rho=1$ was calculated in \pshort.  Recalling
the relation between the $(\rho, \epsilon)$ and $(u,v)$ coordinates,
one finds
		\eqn\neqone{
	{\p a_1\over\p u} = C\cdot\epsilon^{1/2}\Lambda^{-3/2},
	\qquad {\p a_1\over\p v} = C\cdot\epsilon^{-1/2}\Lambda^{-3/2},
		}
for some constant $C$.  So as one approaches the $\bZ_3$
point $\epsilon\rightarrow0$ indeed implies $\p a_1/\p u \rightarrow 0$.

The fact that $\p a_1/\p v \rightarrow \infty$ as $\epsilon^{-1/2}$
implies that the vev of the lowest component of the monopole
hypermultiplet scales as ${M_1/\Lambda} \sim \nu^{1/2}
({\epsilon/\Lambda})^{1/4} $ near the $\bZ_3$ point.  The
$U(1)$ factor with respect to which the light hypermultiplets
are charged has a gap in the $\CN=1$ vacuum which vanishes as
it approaches the $\bZ_3$ point.  Thus the $\CN=1$ $\bZ_3$ vacua
with a ${\rm tr}\phi^3$ superpotential
may be non-trivial fixed points.

\newsec{Physical Interpretations of the $\bZ_3$ Vacua}

First, we remind the reader that we have only the low-energy effective
Lagrangian, and a certain amount of inference will be necessary.
Everywhere except exactly at the $\bZ_3$ point, the massless degrees of
freedom are two $U(1)$ gauge multiplets.  The $\bZ_3$ point itself is
somewhat ambiguous---as we saw, we can keep $\tau$ at any fixed value
as we take the limit $\epsilon\rightarrow 0$.

Near the $\bZ_3$ point, there are two basic scales in the theory.  Most
of the particles have mass $O(\Lambda)$, as is generic in the strong
coupling regime.  In particular, from \chrgs,
all particles with zero magnetic charge
have mass $O(\Lambda)$, which includes the original charged $SU(3)$ gauge
bosons.

The masses of the light hypermultiplets set another
scale.  Generically there are three hypermultiplets with mutually
non-local charges under a single $U(1)$, with comparable masses
$m_i\sim m$.  For $m<<\Lambda$ the kinetic term is almost diagonal and
at scales below $\Lambda$ the other degrees of freedom decouple.
Limits exist in which one of the three hypermultiplets is much lighter
than the other two, e.g. $m_1<<m_2\sim m_3\sim m$, and then the low
energy $U(1)$ coupling depends on $m_1$ in just the way we expect for
$U(1)$ gauge theory containing only that particle at scales below $m$.
There is a complete symmetry between the three hypermultiplets.

Very strikingly, the low energy coupling was independent of the
separation of scales $m/\Lambda$.  We infer that the gauge coupling
does not run below the scale $\Lambda$.  The dependence we did find was
on ratios of particle masses, and the simplest interpretation of this
is that the gauge coupling does not run above the heaviest particle
mass.  Thus we conclude that the theory is essentially an RG fixed
point between scales $\Lambda$ and $m$.  (The qualifier ``essentially''
allows for possible relevant operators associated with the scale $m$.)
No such fixed point exists for $U(1)$ gauge theory with ordinary
charged matter---the mutually non-local charged matter must play an
essential role.

Other interpretations are conceivable if there is additional physics at
an intermediate scale $\mu$, for example the beta function might be
negative above $\mu$ and positive below $\mu$.  Now we know that BPS
saturated states do not appear at intermediate scales, but it is not
clear to us how to disprove the hypothesis that non-BPS states might be
associated with a scale such as the $\epsilon$ of section 4, satisfying
$m\sim\epsilon^{5/2}/\Lambda^{3/2}$.  This would also be a very novel
field theory, and we see no evidence for the additional complications
of this scenario.  Rather than evidence for an intermediate scale, we
will later interpret appearances of $\epsilon$ as the consequence of
non-trivial anomalous dimensions in the fixed point theory.

In the limit $m\rightarrow 0$, the theory becomes a non-trivial RG
fixed point theory.  Although the coupling $\tau$ of the low-energy
theory is adjustable, this is the coupling below the scale $m$, not the
fixed point coupling.  We can find the fixed point coupling by
considering the special case $\rho=0$ where all three particles have
the same mass.  Since they decouple in a symmetric way, the coupling
should be the same above and below this scale.  Thus the fixed point is
at strong coupling, $\tau_c=e^{2\pi i/3}$.

One possibility is that it is a non-trivial fixed point involving
precisely these three particles.  Although each separately would
produce a beta function, their effects cancel out.  Another possibility
is that additional particles our analysis did not detect cancel the
beta function.  Finally, it is conceivable that somehow none of the
three particles contribute to the beta function above the scale $m$.

Let us briefly discuss the last possibility.  It is familiar that
particles stop contributing to the beta function at energies below
their mass, but how can a particle not contribute above an energy $E$?
Now there is a well-known case of a charged `particle' in gauge theory
which does not affect the beta function above an energy $E$.  It is the
conventional 't Hooft-Polyakov monopole, which at weak electric
coupling is generally taken not to affect the beta function at all.
Since its magnetic gauge coupling is large, we should explain why at
higher energies than its mass $m_W/g^2$, it does not produce a one-loop
beta function.

The explanation is familiar.  The monopole has a size of $O(1/m_W)$,
and has small matrix elements with local operators on shorter length scales.
This is the sense in which the monopole decouples above the energy $m_W$.

Could it be that all three particles in the present theory are extended
objects, with sizes (at least) $1/m$, and thus do not contribute to the
beta function above this scale?  We would have a fixed point theory
between the scales $\Lambda$ and $m$ but an effectively trivial one.
Although this is a consistent interpretation of the information given
to us by the effective theory, it does not seem very likely to us.  The
charged particles were monopoles and dyons of the original $SU(3)$
theory and even in the quantum theory, a particle with magnetic charge
must contain some core with unbroken non-abelian symmetry.  One expects
the size of the particle to be $1/\Lambda$, the scale set by
the mass of the $SU(3)$ gauge bosons.
This is clear semiclassically and a non-trivial check
of it at strong coupling was discussed in \DS.

Another way to suppress the beta function would be for the three
particles to form some sort of confined neutral bound state below the
scale $\Lambda$.  However, it is very hard to see how a single charged
particle could be liberated at scales below $m$.  Nevertheless we
mention these possibilities for completeness.

What about the second possibility, that we have not yet identified all
the massless particles?  Perhaps the most interesting version of this
idea begins by noting the similarity of this point with the
``non-abelian Coulomb'' points found in the moduli space of $\CN=1$
supersymmetric gauge theories in \refs{\seione,\SIi}.  The prototypical
example of such a point is the point $\vev{\phi}=0$ in $\CN=4$, $SU(2)$
super Yang-Mills theory, with unbroken non-abelian gauge symmetry.  The
classical central charge formula is valid with $a=\phi$ and
$a_D=\tau\phi$, so both the charged gauge bosons and the monopoles
become massless at this point.  The beta function is zero at all
scales, and the gauge coupling is freely adjustable.

Non-trivial fixed points at unique critical couplings $\tau^*$ are
believed to exist \sei\ in $\CN=1$ supersymmetric QCD with $3N_c/2<
N_f <3N_c$ fundamental flavors, or with an adjoint chiral superfield
and $2N_c/3< N_f <2N_c$ flavors~\kuta.  These theories exhibit
dual descriptions in terms of
(different) light magnetic degrees of freedom.  These degrees of
freedom are presumed to be solitons in the original electric variables
and thus both electrically and magnetically charged light states are
present.
A particularly suggestive analogy can be drawn to $\CN=1$, $SO(3)$ gauge
theory with two flavors~\refs{\SIi,\SIii}
which can be described using fundamental
electron, monopole, or dyonic variables.\footnote*{
We thank N. Seiberg for this example and the argument below.}

The non-trivial fixed points are perhaps more similar to what we found,
but the main point seems to be that all of these fixed point theories
contain non-Abelian gauge bosons
which produce a negative contribution to the beta function.

The known models also contain additional matter---at
least two doublet hypermultiplets
to realize the model of \kuta.
This already leads to severe difficulties with this identification,
as it implies that the model has
additional global symmetry, and additional
`Higgs branches' of the moduli space reached by turning on
large squark vevs, all of which should have been visible in the semiclassical
limit.$^*$  One might still imagine that we have discovered
a non-abelian Coulomb phase without additional matter.

We now argue that there are no massless
non-abelian gauge bosons at the $\bZ_3$ point.
Now the non-abelian gauge bosons as well as the additional matter we would need
are all BPS saturated states.  By moving away from the $\bZ_3$ point in the
$\CN=2$ moduli space, we can make the generic BPS state massive.
On the other hand, we will see that for every possible charged state, there
is a line we can move on which keeps only it (and states with multiples of its
charge) massless.  The resulting low energy theory is a standard $U(1)$ gauge
theory with mutually local charges, and the charged spectrum will be reflected
in the low energy gauge coupling by standard field theoretic considerations.
Thus, from the effective Lagrangian we can make strong statements about the
existence of charged BPS particles.

The minimal test for a candidate `$W$ boson' is that it have charge $2$
in some basis.  If we were to find a candidate, we would then need to check
that it is in a vector multiplet, but this will not be necessary.
There are two possibilities;
in the basis in which the three known hypermultiplets
have charge $(1\ 0)$, $(0\ 1)$ and $(-1\ -1)$, the first possibility
is the bound state of two monopoles $(2\ 0)$ (or its $\bZ_3$ images),
while the second possibility
is a more complicated bound state, for example $(2\ 4)$.

Consider the first possibility, e.g. the state $(2\ 0)$.
This is a particle which is believed not to exist semiclassically;
certainly for $SU(2)$ the multi-monopole moduli space is
known to be connected \atiyah\ and there is no distinct charge $|h|>1$
monopole solution, nor is there such a state in the quantized
two-monopole system.
However we do not know a proof for the $SU(3)$ theory at hand,
and we do not want to rely on semiclassical results, so we will
argue as follows.
If the particle existed near the $\bZ_3$ point, we could follow it
out along a trajectory near the line on which the $(1\ 0)$ state was massless,
and it would give an additional logarithmic contribution to the beta function
everywhere along that line, which is not present.
Thus this particle does not exist.
Indeed we could follow the line all the way to a $\bZ_2$ point, with no signal
from the effective Lagrangian of any instability, so the existence of
the particle would probably invalidate the rather
successful physical picture of the $\CN=1$ theory given in
\refs{\SWi,\AF,\DS}.

The other possibility is a state like $(2\ 4)$.
If it were possible to find a state which became massless
only at the $\bZ_3$ point itself, it would be available to cancel the beta
function between $\Lambda$ and $m$, but the previous argument would not apply.

This is not possible.  A state with charges $(m\ n)$ will become massless
along the line $m a_1+n a_D^1=0$ or equivalently $a_D^1/a_1=-m/n$.
In section 4, we saw that there exist trajectories out of the $\bZ_3$ point on
which $a_D^1/a_1$ takes any fixed real value, with fixed $\rho$ and $\epsilon$
variable.  Along this line, the massless charged spectrum
of the low energy theory is again local and if these states existed they
would have left a direct signature in the effective Lagrangian.  We
conclude that only the three hypermultiplets we already identified exist.

This argument from the singularities of the effective Lagrangian is quite
general.  For example, it applies to the $SU(2)$ solutions of \SWii\ and shows
that for $N_f\le 2$, no BPS state with magnetic charge $h\ge 2$ is stable
anywhere in moduli space, while for $N_f=3$, only the $(2,1)$ state and its
images $(2,2k+1)$ exist.

\newsec{$U(1)$ Gauge Theory with Mutually Non-local Charges}

Both the evidence from the effective Lagrangian, and consideration
of alternate interpretations, led us to conclude
that the low energy theory of the $\bZ_3$ point is a field theory of a truly
novel type, a non-trivial interacting $\CN=2$ superconformal
theory involving two $U(1)$ gauge multiplets and
three hypermultiplets with mutually non-local charges under one of the
$U(1)$'s.
We now take the limit $\Lambda\rightarrow \infty$, after which only these
degrees of freedom remain.
This is the degeneration limit of the quantum surface and
the $U(1)^2$ effective Lagrangian was described in section 4.

The resulting theory still has a two complex dimensional space of vacua.
Generically, all the hypermultiplets are massive, and there should be no
difficulty in making a particle interpretation of the theory.  At energies
where pair creation is unimportant, the known quantum mechanical
formulations
(for example \yang) should apply.
By tuning the mass of one hypermultiplet to zero, one produces massless
QED
with additional massive monopoles and dyons, which again should not be
much more difficult to treat than massless QED.

By tuning to the $\bZ_3$ point,
or adding the superpotential $\Tr\phi^3$,
we produce a superconformal theory.
Whether a particle interpretation can be made for it is not at all clear.
The effective Lagrangian gives only limited insight into the physics,
because the fixed point is strongly coupled.
Clearly an important problem for future work is to find a more direct
treatment of this theory.
The following are some observations and speculations in this direction.

Let us try to write a bare Lagrangian at the scale
$\Lambda$ which defines the fixed point theory and its relevant perturbations.
It will certainly contain the hypermultiplets and the non-trivial $U(1)$.

It is not as obvious whether it must contain the second
$U(1)$ under which the hypermultiplets are neutral.
In the $\CN=1$ theory with a superpotential, the vacuum is determined as
discussed in section 4, and it would appear that one requires both flat
directions and thus both $U(1)$'s to describe this.
In the $\CN=2$ theory the second $U(1)$ decouples in the following sense.
First, its vector potential and fermions might couple through
off-diagonal gauge kinetic terms.
{}From \split, the corresponding terms in the
effective Lagrangian were higher dimension operators with
the standard suppression by inverse
powers of the cutoff, $\sim (a_1/\Lambda) W^1 W^2$, and it is very plausible
that this is also true at the scale $\Lambda$, so that the
vector potential and fermions decouple as $\Lambda\rightarrow \infty$.

There is also explicit dependence of $\tau^{11}$ on the scalar $a_2$.
{}From \cflim,
$\tau^{11}$ depends on $a_2$ as a function of
$g_2^{5/4}/a_1\Lambda^{3/2} \sim (a_2-\hat{a_2})^{5/4}/a_1\Lambda^{1/4}$
and at the scale $m\sim a_1$ these couplings are suppressed by powers of
$((a_2-\hat{a_2})/\Lambda)^{1/4} \sim (a_1/\Lambda)^{1/5}$.
Essentially, the leading operator coupling the two sectors
in the effective Lagrangian,
$((a_2-\hat{a_2})/\Lambda)^{1/4}(\p a_1)^2$, is irrelevant.

Thus it appears that we can drop the second $U(1)$, but we still need
to allow varying the expectation value of $a_2$.
We could reproduce this if the fixed point theory
has a relevant perturbation $(a_2-\hat{a_2})~O_2$
producing the flow to $\CN=2$ supersymmetric QED.
This would make sense if $O_2$ has dimension $\Delta>3$, so that the
expectation value $\vev{a_2-\hat{a_2}}O_2$ is relevant but the fluctuation
$\delta a_2 O_2$ is irrelevant.
(Such an operator, irrelevant in the UV limit of
a flow but becoming relevant in
the IR, is referred to as a `dangerous irrelevant operator' in critical
phenomena.)
The dependence of the low-energy Lagrangian on $(a_2-\hat{a_2})^{5/4}$
would be explained if $O_2$ had dimension $\Delta=3+1/5$.

A similar discussion can be made for the $N$-fold critical point present
in $SU(N)$ gauge theory, and leads to the conclusion that the decoupled
$U(1)$'s are associated with a series of irrelevant operators
with dimension $5-2n/(N+2)$, $2\le n< N/2+1$.

To couple to all three hypermultiplets, we need a Lagrangian including both the
gauge field and its dual.  Such a Lagrangian has been written by Schwarz and
Sen \ss; however Lorentz invariance and
$\CN=2$ supersymmetry are not manifest.
Let us assume such Lagrangians exist but for now not use specific forms
or symmetry properties.
Indeed, we do not know a priori what is an appropriate gauge Lagrangian
for this strongly coupled theory.
Below the scale $m$,
the gauge field fluctuations are of course controlled by the usual quadratic
kinetic term, but to emphasize our uncertainty about this point in the fixed
point theory, we allow an additional unknown term $S_0$ in the action.

We write
		\eqn\lagr{\eqalign{
S &= S_{0}(V,V_D,A,A_D) + {\tau\over 4\pi}S_{free}(V,V_D,A,A_D)
+ (\vev{a_2}-\hat{a_2})O_2\cr
&+\int d^4x d^4\theta\ E^+ e^V E + M^+ e^{V_D} M + D^+ e^{V+V_D} D
+ {\rm charge\ conj.}\cr
&+ \sqrt{2}~
 \int d^4x d^2\theta\ E A \tilde E + M A_D\tilde M + D(A+A_D)\tilde D +
 {\rm h.c.} .
		}}
The fields $A_D$ and $A$ are not independent.
We know the relation $A_D=\p\CF(A,A_2)/\p A$ in the effective Lagrangian,
which we interpret by replacing $A_2$ with the vev of its scalar component.
One test of the bare Lagrangian would be to reproduce this relation
in the low energy limit.

Given a particular $S_0$, or even assuming $S_0=0$,
a serious obstacle to using this directly is that the fixed point is
at strong coupling.  Nevertheless, let us see what we can do.
The simplest check would be to find a zero of the beta function.
There is a simple ansatz which produces the correct fixed point coupling,
illustrates what is going on, and might be correct.  It is to
compute the contribution of each hypermultiplet separately to the beta
function of its `natural' gauge coupling, and then use $SL(2,\bZ)$ to
express the answers in terms of a single coupling.

By $\CN=2$ supersymmetry, the only perturbative contribution to the
beta function will be at one loop.  In $D=4$ $U(1)$ gauge theory with a
single hypermultiplet, one does not expect non-perturbative contributions.
This is the standard one-loop renormalization
                \eqn\betadef{
        \Delta\CL_{eff} = {i\over\pi}g_i^2 \log \Lambda_0^2~
        \left(\int d^4\theta\ A_i A_i^+ +
        \int d^2\theta\ W_i^2 + c.c. \right)
                }
which does not depend on the (unknown) gauge Lagrangian.

Let us then take as the beta function due to the electron
		\eqn\elbeta{
	{\p\over\p\log\mu}\tau = -{i\over 2\pi}.
		}

A theory with a single hypermultiplet with charge $(q_m,q_e)$
is just as easy to deal with,
by using $SL(2,\bZ)$ to write the gauge action in terms of
$V_{(q_m,q_e)}=q_m V_D + q_e V$.
This will transform the original electric $\tau$ into
		\eqn\newtau{
	\tau_{(q_m,q_e)}={1\over \gcd(q_e,q_m)}{a\tau+b\over q_m\tau+q_e}
		}
with $a$ and $b$ integers chosen to make $a q_e-b q_m=\gcd(q_e,q_m)$.
Transforming back, it contributes the electric beta function
		\eqn\elbeta{\eqalign{
	{\p\over\p\log\mu}\tau &=
	{\p\tau_{(0,1)}\over\p\tau_{(q_m,q_e)}}{\p\over\p\log\mu}
	\tau_{(q_m,q_e)}\cr
	&= -{i\over 2\pi} (q_m\tau+q_e)^2.
		}}

At this point the main assumption is duality in the quantum theory.
In general, magnetically charged hypermultiplets make the electric coupling
relevant, as was observed in the context of the dual $U(1)$ Lagrangian of \SWi.
An interesting feature of the result is that with magnetically charged
hypermultiplets, the flow depends on and can change the real part of $\tau$.
Although normally $U(1)$ gauge theory in $D=4$ is unaffected by the
$\theta$ angle, theories containing electric and magnetic charges can
be affected.

A very natural ansatz for the total beta function is simply to add the
individual electric beta functions.  The main justification
we will give for this
is the observation that the duality transforms of \betadef\ are
expressed in terms of $W$, $W_D$, $A$ and $A_D$, which are all
locally related operators, unlike the vector potentials.  Thus, despite
the subtleties associated with mutually non-local charges,
we can make sense of the RG.

We thus find the condition for a fixed point:
		\eqn\fixed{
	\sum_i (q_{mi}\tau_c+q_{ei})^2 = 0.
		}
For our spectrum we have $1+\tau_c^2+(\tau_c+1)^2=0$ implying
		\eqn\ourpoint{
	\tau_c=e^{2\pi i/3},
		}
so this simple ansatz produces the correct fixed point coupling.
Now this is not really convincing evidence that the beta function
is exact; it might be that the corrections also cancel.
Indeed, the $\bZ_3$ symmetry of the spectrum essentially
guarantees that there will be a fixed point at this $\tau_c$.
The beta function will satisfy $\beta(g(\tau))=g(\beta(\tau))$
for any symmetry $g\in SL(2,\bZ)$ of the quantum theory, and thus fixed
points of the symmetry will be fixed points of the RG.

Another interesting property of this beta function is that its
fixed point is attracting.
There is an even simpler fixed point theory which we will use
to demonstrate this---the theory containing only an
electron and a monopole.  Now since we do not
have a realization of it, we cannot be sure that this theory is
consistent.  Perhaps consistency conditions not yet known to us
require three hypermultiplets.  In any case, this calculation
of its beta function does make sense, and produces $\tau_c=i$.
Then, expanding
$\tau=i+\delta$, for small $\delta$ the flow becomes
$d\delta/ds = -\delta/\pi$.  The flow in the three particle model
around $\tau_{c3}$ is similar, with $d\delta/ds =
-(\sqrt{3}/\pi)\delta$.

The meaning of the result is that $S_{free}$ is
an irrelevant operator.  If we believe the one-loop result is exact,
we find its anomalous dimension (in the three-particle theory)
to be $\sqrt{3}/\pi$.
Now we already saw a non-trivial dimension for the operator $O_2$,
which we inferred from effective field theory results such as \cflim,
and it was a rational number.
We consider the non-rationality of the anomalous dimension we found for
$S_{free}$ to be some evidence against its exactness -- perhaps
non-perturbative effects are present, or perhaps $S_0$ is important.

Whether or not this anomalous dimension is exact,
it is very suggestive that it is positive.
In a unitary conformal theory, the anomalous dimension of a scalar field
is required to be non-negative, and zero anomalous dimension implies that
it
is a free field \CFT.  Thus this result is consistent with the non-trivial
nature of the fixed point.

Since this is a superconformal theory, there must be an
unbroken $U(1)_R$ symmetry, and unlike the models
of \refs{\sei,\kuta,\kutsch}\ there is no obvious candidate
for this among the original symmetries.
A novel feature following from the generation of massless solitons,
is the generation of new chiral symmetries.
Indeed, it would appear that each massless hypermultiplet will come with
its own $U(1)_L\times U(1)_R$ fermion number symmetries.
We expect the chiral symmetry to be anomalous, and duality transforming
the
standard chiral anomaly for electrically charged hypermultiplets,
the monopole also has the standard anomaly, but dyons will have terms such
as
\eqn\weird{\partial^\mu J_{\mu5} = \ldots + q_e\cdot q_m (E^2-B^2).}
Although it may seem difficult to cancel anomalies such as \weird,
we have already seen it implicitly in the statement that summing
the flows \betadef\ produces a fixed point.

A toy model for some of these phenomena would be an analogous theory
in two dimensions, with matter couplings to a scalar field $\phi$ and its
dual $d\phi = *d\tilde\phi$ producing a beta function for each.  Of course
the dimensional reduction of the present theory would be such a model, but
simpler possibilities might exist.
It would be interesting to study the topologically twisted form of
the $D=4$ theory as well.

\newsec{Conclusions}

We have given strong evidence for the existence of a sensible,
$D=4$, $\CN=2$ supersymmetric $U(1)$ gauge theory containing an
electron, a monopole and a dyon hypermultiplet, as a special vacuum of
$SU(3)$ gauge theory.  We believe this is the first strong evidence
that Dirac's original conception of gauge theory containing fundamental
electrons and monopoles can be realized in a fully consistent local
relativistic theory.

Furthermore, the theory at the $\bZ_3$ point
may well be the simplest non-trivial $D=4$, $\CN=2$ superconformal theory.
We proposed a general mechanism to produce fixed points in theories with
mutually non-local charges, and saw that it fit the data from the effective
theory.  We identified non-trivial critical exponents in the theory.

Clearly the main problem for future research is to construct and work
with a direct definition of the theory.  There is a series of
generalizations at special vacua of $SU(N)$ gauge theory to study as
well.  We have no doubt that this will shed much light on $D=4$
superconformal field theory, and field theory in general.

\bigskip
\centerline{{\bf Acknowledgements}}

It is a pleasure to thank A. Faraggi, D. Friedan,
K. Intriligator, J. March--Russell,
R. Plesser, N. Seiberg, A. Shapere, S. Shenker,
M. Strassler and E. Witten for useful
conversations.
The work of P.C.A. is supported in part by NSF grant PHY92-45317 and by the
Ambrose Monell Foundation, while the work of M.R.D. is supported in
part by DOE grant DE-FG05-90ER40559, NSF
PHY-9157016 and the A. P. Sloan Foundation.

\appendix{A}{Calculation with Elliptic Functions}

The relation between $(g_2,g_3)$ and $(\omega,\omega_D)$ is standard \Ell.
A convenient normalization for the $(g_2,g_3)$ functions is
		\eqn\geisen{
	g_2 = {4\pi^4\over 3\omega^4}E_4(\tau), \qquad
	g_3 = {8\pi^6\over 27\omega^6}E_6(\tau) ,
		}
with $q=e^{2\pi i\tau}$.  These are modular functions in $\tau$
satisfying $E_k(-1/\tau)=\tau^k E_k(\tau) + \delta_{k,2} (12/2\pi i)\tau.$

The $a_i$ can be calculated by using $x=\wp(\nu)$ and $w=\wp'(\nu)$ to
write
		\eqn\lamrew{
	\int\tilde\lambda = \int d\nu~(\wp'(\nu))^2
	= {1\over 30}\wp''' + {2\over 5}g_2\zeta - {3\over 5}g_3\nu
		}
(from standard tables or simply by differentiating the result and
using \branchtwob).
The function $\zeta=-\int\wp$ is not single valued and we have
		\eqn\zetarel{\eqalign{
	\zeta(\nu+\omega)-\zeta(\nu) &\equiv 2\eta =
	{\pi^2\over 3\omega} E_2(\tau) \cr
	\zeta(\nu+\omega_D)-\zeta(\nu) &\equiv 2\eta_D =
	{\pi^2\over 3\omega_D} E_2(-{1\over\tau})
	 = {\pi^2\over 3\omega} \tau~E_2(\tau) - {2\pi i\over\omega} \cr
		}}
where $E_2(\tau) = 1 - 24q + \ldots$.  This implies
		\eqn\etaega{
	\eta\omega_D-\eta_D\omega = i\pi.
		}
Combining these results,
		\eqn\lamperi{\eqalign{
	a_1 &\equiv \int_0^{\omega} \tilde\lambda \cr
	&={8\pi^6\over45\omega^5} \left(E_2(\tau)~E_4(\tau)-E_6(\tau)\right)
	= -i{4\pi^5\over15\omega^5} {\p\over\p\tau}E_4(\tau) \cr
	a_D^1 &\equiv \int_0^{\omega_D} \tilde\lambda \cr
	&= {8\pi^6\over45\omega_D^5}
	\left(E_2(-{1\over\tau})~E_4(-{1\over\tau})-E_6(-{1\over\tau})\right)
	= \tau a_1 - i{16\pi^5\over 15 \omega^5} E_4(\tau) \cr
	&= \tau a_1 - i{4\pi\over 5 \omega} g_2 ,
		}}
implying \nicerel.

\bigskip
\listrefs
\end